\documentclass{aa}  
\usepackage{graphicx,amssymb,amsmath,hyperref,multirow}
\usepackage[usenames,dvipsnames]{xcolor}

\usepackage{rotating}
\usepackage{txfonts}
\usepackage{lscape}
\usepackage{amsmath}
\usepackage{natbib,twoopt}
\usepackage{longtable}
\usepackage{multicol}
\bibpunct{(}{)}{;}{a}{}{,} 
\newcommandtwoopt{\citeyearads}[3][][]%
{\href{http://adsabs.harvard.edu/abs/#3}{\citeyear[#1][#2]{#3}}}
\hypersetup{colorlinks,linkcolor=red,citecolor=blue,urlcolor=blue}

\newcommand{\feh}{[Fe/H]}
\newcommand{\teff}{$\mathrm{T}_{\mathrm{eff}}$}
\newcommand{\logg}{$\log g$}
\newcommand{\vmic}{$v_{\mathrm{mic}}$}
\newcommand{\vmac}{$v_{\mathrm{mac}}$}
\newcommand{\vsini}{$v{\sin{i}}$}
\newcommand{\kms}{km~s$^{-1}$}

\newcommand{\cygA}{\object{61~Cyg~A}}

\newcommand{\tab}[1]{Table~\ref{#1}}
\newcommand{\fig}[1]{Fig.~\ref{#1}}
\newcommand{\sect}[1]{Sect.~\ref{#1}}

\newcommand{\hfsn}{HFS:N}
\newcommand{\hfsy}{HFS:Y}

\newcommand{\porto}{{\tt PORTO}}
\newcommand{\ucm}{{\tt UCM}}
\newcommand{\sme}{{\tt SME}}
\newcommand{\ispec}{{\tt iSpec}}
\newcommand{\bol}{{\tt BOL}}
\newcommand{\ulb}{{\tt ULB}}

\newcommand{\kdwarf}{$K$-dwarfs}
\newcommand{\fgkgiant}{$FGK$-giants}
\newcommand{\fgdwarf}{$FG$-dwarfs}
\newcommand{\metalpoor}{metal-poor}

\begin{document}

   \title{Gaia FGK Benchmark stars: Opening the black box of stellar element abundance determination}


   \author{P.~Jofr\'e\inst{\ref{ioa},\ref{udp}}
   \and U.~Heiter\inst{\ref{uppsala}}   
   \and C.~C.~Worley\inst{\ref{ioa}}
   \and S.~Blanco-Cuaresma\inst{\ref{geneva}}
    \and C.~Soubiran\inst{\ref{bx}}
   \and T.~Masseron\inst{\ref{ioa}}
   \and K.~Hawkins\inst{\ref{ioa}, \ref{col}}
   \and V.~Adibekyan\inst{\ref{porto}}
   \and S.~Buder\inst{\ref{mpia}}
   \and L.~Casamiquela\inst{\ref{barcelona}}
   \and G.~Gilmore\inst{\ref{ioa}}
   \and A.~Hourihane\inst{\ref{ioa}}
   \and H.~Tabernero\inst{\ref{UCM}}
    }

\authorrunning{Jofr\'e et al. }
\titlerunning{Benchmark stars: systematic uncertainties in abundance determination}
\offprints{ \\ 
P. Jofr\'e, \email{pjofre@ast.cam.ac.uk}
}

   \institute{Institute of Astronomy, University of Cambridge, Madingley Road, Cambridge CB3 0HA, United Kingdom \label{ioa}
         \and N\'ucleo de Astronom\'ia, Facultad de Ingenier\'ia, Universidad Diego Portales,  Av. Ej\'ercito 441, Santiago, Chile \label{udp}
         \and Department of Physics and Astronomy,  Uppsala University, Box 516, 75120 Uppsala, Sweden \label{uppsala}
         \and  {Observatoire de Gen\`eve, Universit\'e de Gen\`eve, CH-1290 Versoix, Switzerland}\label{geneva}
         \and Laboratoire d'Astrophysique de Bordeaux, Univ. Bordeaux, CNRS,UMR 5804, F-33615, Pessac, France\label{bx}
         \and Department of Astronomy, Columbia University, 550 W 120th St, New York, NY 10027\label{col}
	  \and {Instituto de Astrof\'isica e Ci\^encias do Espa\c{c}o, Universidade do Porto, CAUP, Rua das Estrelas, 4150-762 Porto, Portugal} \label{porto}
         \and {Max-Planck Institute for Astronomy, 69117, Heidelberg, Germany}\label{mpia}
         \and Departament d'Astronomia i Meteorologia, Universitat de Barcelona, ICC/IEEC, E-08007 Barcelona, Spain \label{barcelona}
	   \and {Dpto. Astrof\' isica, Facultad de CC. F\' isicas, Universidad Complutense de Madrid, E-28040 Madrid, Spain} \label{UCM}   
             }

   \date{Received XXXX, 2016; accepted XXX, 2016}

 
  \abstract
   {Gaia and its complementary spectroscopic surveys combined will yield the most comprehensive database of kinematic and chemical information of stars in the Milky Way. The Gaia FGK benchmark stars  play a central role in this matter as they are calibration pillars for the atmospheric parameters and chemical abundances  for various surveys.  The spectroscopic  analyses of the benchmark stars are done by combining different methods, { and the results will be affected by the systematic uncertainties inherent in each method}. In this paper we explore some of these systematic uncertainties. We determined { line} abundances of Ca, Cr, Mn and Co for four benchmark stars using six different methods. { We changed the default input parameters of the different codes in a systematic way and found in some cases significant differences between the results.} { Since there is no consensus on the correct values } for many of these default parameters, we urge the community to raise discussions towards standard input parameters that could alleviate the difference in abundances obtained by different methods. { In this work we provide quantitative estimates of uncertainties in elemental abundances due to the effect of differing technical assumptions in spectrum modelling.} }


   \maketitle
%

\section{Introduction}

In the times of the first data release of Gaia, Galactic astronomy is experiencing a revolution due to the millions of spectra of individual Milky Way stars becoming available to the community thanks to the several on-going  and future high-resolution spectroscopic surveys. Some prominent ones are APOGEE \citep{2014ApJS..211...17A}, Gaia-ESO \citep[hereafter GES]{2012Msngr.147...25G}, GALAH \citep{2015MNRAS.449.2604D},  4MOST \citep{2012SPIE.8446E..0TD} and WEAVE \citep{2012SPIE.8446E..0PD}. Because these datasets have different selection functions attempting to survey stars of different nature, the spectra are different from each other and therefore different abundance analysis pipelines are developed independently, each affected by their own systematic uncertainties. In order to extract the best information out of these rich datasets, it became important to  combine the complementary data from these surveys and therefore the attempts of cross-calibrating surveys became a principal activity in the field \citep{2015arXiv150608642F}. Cross-calibration can best be done when considering stars observed in common between surveys \citep{2016arXiv160207702B, 2016arXiv160200303H}.  

The Gaia FGK Benchmark Stars (GBS) \citep[][hereafter Paper\,I]{Heiter2015a} have become a crucial resource in the cross-calibration of stellar parameterisation methods and stellar population surveys. Provided as a spectral library \citep[][hereafter Paper\,II]{BlancoCuaresma2014} with which different instruments and resolutions can be simulated, key studies have extended the GBS known characteristics from stellar parameters ($T_{\textrm{eff}}$ and $\log g$, Paper\,I) to metallicity \citep[][hereafter Paper\,III]{Jofre2014}, $\alpha$ and Fe--peak element abundances \citep[][hereafter Paper\,IV]{Jofre2015} and extensions of the sample towards low metallicities \citep[][Paper\,V]{2016A&A...592A..70H}. See also \cite{2016AN....337..859J} for summary. 

The GBS spectra are of exquisite quality, with excellent signal-to-noise (S/N) ratio and very high resolution. Thus their analysis has provided the community with key reference parameters and element abundances. The GBS are being considered as calibrators by GES \citep{2013Msngr.154...47R}, GALAH \citep{2015MNRAS.449.2604D} and RAVE \citep{2016arXiv160903210K}. APOGEE  also has observed some of them in order to evaluate results.  In addition, some independent works on spectral analyses are using the GBS as a tool to assess the quality of results. Examples are the AMBRE project \citep[e.g.][]{2016A&A...591A..81W}, or the introduction of new methods \citep{2014MNRAS.443..698S, 2016A&A...587A...2B, 2016arXiv160408800H, 2016arXiv160808392T}.

\begin{table*}[t]
\small
\caption{Atmospheric parameters of the four GBS from Papers\,I and III. \vmac\ was used only as input parameter by  \sme\ and \ispec. }\label{t:bs_ap}
\begin{tabular}{lcccccc}
\hline\hline
Star & $T_{\textrm{eff}}$  & $\log g$ & [Fe/H] & \vmic & \vsini  & \vmac\\
        &    K   &  dex &  dex    & \kms & \kms & \kms\\
\hline 
Sun & 5771$\pm$01 & 4.44$\pm$0.00 & 0.03$\pm$0.05& 1.06$\pm$0.18 & 1.6 & 4.2 \\
Arcturus & 4286$\pm$35 &1.64$\pm$0.09 &-0.52$\pm$0.08 &1.58$\pm$0.12 &3.8  & 5.0\\
61CygA & 4374$\pm$22 & 4.63$\pm$0.04  &-0.33$\pm$0.38  &1.07$\pm$0.04 & 0.0 & 4.2\\
HD22879 &  5868$\pm$89 & 4.27$\pm$0.03 &-0.86$\pm$0.05& 1.05$\pm$0.19 & 4.4 &5.4\\
\hline
\end{tabular}
\end{table*}

In the spirit of avoiding systematic uncertainties due to a given method, the results of the spectral properties of the GBS come from combining several different methods. This follows the spirit of the GES parametrisation strategy \citep{2014A&A...570A.122S}.  Although the expectation is that this strategy yields more precise results, the combination of different methods is not trivial as discrepancies between results can be large and usually very difficult to improve. While  discussions of this problem are found thanks to GES efforts in \cite{2014A&A...570A.122S} and in Paper~III and IV, as well as in \cite{2014AJ....148...54H} and \cite{2016arXiv160408800H}, other dedicated works in this direction have also been reported. The latter come from massive efforts  concluding from workshops, where experts met and analysed common stars with different methods \citep{2012A&A...547A.108L, 2016arXiv160703130H}.

In particular, in Paper~IV we found that differences in spectral analyses methods can lead to differences in abundances of more than 0.3~dex, even when stellar parameters and atomic data were kept fixed.  To overcome  this problem in Paper~IV we combined the abundances in a differential approach. 
Differential analyses with respect to the Sun have shown to significantly improve the accuracy of results \citep[e.g.][]{2014A&A...562A..71B, 2015A&A...579A..52N, 2015MNRAS.453.1428J, 2016A&A...590A..32T}. Since the GBS are very different from each other, in Paper IV they were separated in groups and the abundances were derived differentially taking one reference star within each group. The reference stars were then carefully analysed differentially with respect to the Sun. 

The differential approach, however, does not solve the issue of having sometimes disagreements between methods that cannot be justified given the same material and input parameters used for the analysis, and does not provide clues to the reason of these discrepancies.  Thus,  with the team determining abundances of GBS we proposed to investigate what  might  cause  these differences. The aim is essentially to ``open the black box'' of each pipeline and investigate which of the many hidden assumptions in spectral analysis are responsible of sometimes large abundance differences. { Since for many of the technical assumptions in such analyses there is no consensus on the correct approach, this study also  aims to provide guidelines for  a realistic error estimation.} 

In February 2016 we organised a 4-day workshop at the Institute of Astronomy in Cambridge, UK, which was attended by one member each representing six out of the eight methods that have been used throughout the GBS work series.   Among these methods, three  were also part of the workshop summarised in \cite{2016arXiv160703130H}. In this article we report our findings.  We describe first the data we used in \sect{data}, in \sect{methods} we describe our methods, in \sect{results} we discuss each test and present the results. We finalise with a summary and concluding remarks in \sect{concl}.

\begin{table}[t]
\caption{Atomic line data of the four chosen lines { and the additional Ca line used for the blending test (\sect{blends})} published in Paper\,IV as well as the HFS for \ion{Co}{I} and \ion{Mn}{I}. For Mn, we used the modified line list (see \sect{1.linelist}). {{ Notes}: Ca: all data from \citet{SR}; Cr: wavelength and energies from \citet{WLHK}, loggf from \citet{SLS}; Mn: wavelength, energies, loggf from \citet{DLSSC}, HFS constants for lower level from \citet{1969PhLA...29..486H}, for upper level from \citet{1987ZPhyD...7..161B}; Co: wavelength and energies from \citet{K08}, loggf from \citet{1982ApJ...260..395C}, HFS constants from \citet{1996ApJS..107..811P}}} \label{tab:linelist}
\begin{tabular}{lccccccc}
\hline \hline
Element &  Ion & $\lambda$ &  Ex.Pot & $\log gf$ \\
Ca &      1& 5260.3870 &     2.521 &    -1.719 \\
Ca & 1 & 6455.598 &  2.523 & -1.290\\
Cr &      1& 5238.9610 &     2.709 &    -1.270 \\
Mn &      1& 6021.7933 &     3.075 &    -0.054 \\
Co &      1& 5352.0397 &     3.576 &     0.060 \\
\hline
\multicolumn{5}{l}{HFS}  \\
Mn &      1& 6021.7210 &     3.075 &    -2.756 \\
Mn &      1& 6021.7470 &     3.075 &    -1.539 \\
Mn &      1& 6021.7490 &     3.075 &    -2.404 \\
Mn &      1& 6021.7690 &     3.075 &    -1.363 \\
Mn &      1& 6021.7720 &     3.075 &    -2.279 \\
Mn &      1& 6021.7780 &     3.075 &    -0.621 \\
Mn &      1& 6021.7880 &     3.075 &    -1.337 \\
Mn &      1& 6021.7910 &     3.075 &    -2.358 \\
Mn &      1& 6021.7950 &     3.075 &    -0.761 \\
Mn &      1& 6021.8020 &     3.075 &    -1.404 \\
Mn &      1& 6021.8080 &     3.075 &    -0.919 \\
Mn &      1& 6021.8110 &     3.075 &    -1.580 \\
Mn &      1& 6021.8170 &     3.075 &    -1.103 \\
Mn &      1& 6021.8210 &     3.075 &    -1.325 \\
Mn &      1& { 6021.8210} &     3.075 &    { -1.610} \\
Co &      1& 5351.8931 &     3.576 &    -3.244 \\
Co &      1& 5351.9244 &     3.576 &    -2.855 \\
Co &      1& 5351.9526 &     3.576 &    -1.736 \\
Co &      1& 5351.9770 &     3.576 &    -1.532 \\
Co &      1& 5351.9985 &     3.576 &    -1.459 \\
Co &      1& 5352.0191 &     3.576 &    -0.581 \\
Co &      1& 5352.0321 &     3.576 &    -0.662 \\
Co &      1& 5352.0509 &     3.576 &    -0.503 \\
Co &      1& 5352.0696 &     3.576 &    -0.569 \\
\hline  
\end{tabular}
\end{table}

\begin{figure*}[!ht]
\includegraphics[scale=0.9]{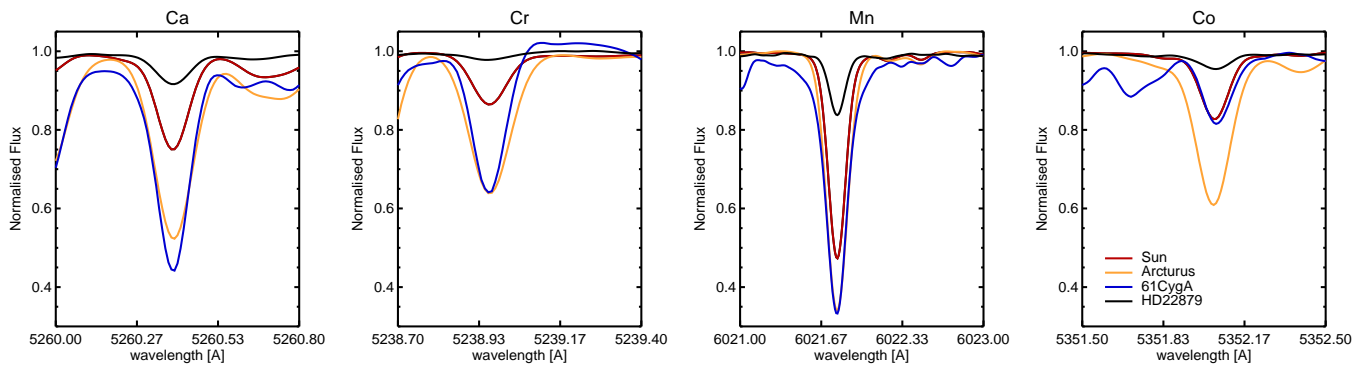}
\caption{Observed profiles of the four different lines that were selected and analysed, for each of the four GBS.}
\label{fig:profiles}
\end{figure*}

\section{Input data}\label{data}
\subsection{Stellar sample}
Four GBS were chosen for  this investigation. They are representative of the groups \fgdwarf, \kdwarf, \fgkgiant\ and \metalpoor: Sun, 61CygA, Arcturus, HD22879 and are used as reference in Paper~IV for the  of the abundance determination. Table~\ref{t:bs_ap} summarises their fundamental \teff\ and  \logg\ from Paper\,I and recommended (NLTE) \feh\ determined in Paper~III as well as the adopted \vmic\ and \vsini. Their high S/N, high resolution spectra, were taken from the GBS library described in Paper\,II. These spectra were previously used for the study of  GBS abundances as determined by several methods with common inputs (Papers\,III and IV).  In particular, the spectra used  here were selected from Paper~IV and come from NARVAL  for 61CygA, { the spectrum of \cite{2005ASPC..336..321H}} for the Sun and Arcturus, and HARPS for HD22879, and were normalised, convolved to a common resolution of 70,000 and corrected by radial velocity (RV) as described in Paper ~II (see also Paper~IV for details).

\normalsize

\begin{table*}[t]
\tabcolsep=0.13cm
\caption{Absolute abundances, A(X), and equivalent widths, EW, of the four chosen lines in each GBS from Paper IV, for the methods considered in this work. }
\begin{tabular}{|c|ccc|cccc|ccccc|}
\hline\hline
Star & \multicolumn{ 2}{c}{Element} & Line & EW &   &  &  & A(X)  &  &  &  &  \\ 
 & Symbol & Z & [\AA] & \porto\ & \bol & \ucm & \ulb & \ispec & \porto  & \bol & \ucm & \ulb \\ 
\hline
Sun & Ca & 20 & 5260.38 & 34.30  & 30.30 & - & 34.31 & 6.33  & 6.25 & 6.23 & - & 6.31 \\ 
 & Cr & 24 & 5238.96 & 15.30  & 14.00 & 15.50 & 16.84 & 5.60  & - & 5.43 & 5.47 & 5.52 \\ 
 & Mn & 25 & 6021.81 & 90.60  & - & - & 59.38 & 5.74 & 5.49 & - & - & 5.38 \\ 
 & Co & 27 & 5352.04 & 24.90 & 21.40 & 24.20 & - & 4.85  & 4.76 & 4.78 & 4.81 & 4.82 \\ 
\hline
Arcturus & Ca & 20 & 5260.38 & 74.20  &70.40 & - & 80.27 & 5.93  & 5.72 & 5.70 & - & 5.87 \\ 
 & Cr & 24 & 5238.96 & 49.90 & - & 50.60 & 51.79 & 5.01 & 4.83  & - & 4.84 & 4.87 \\ 
 & Mn & 25 & 6021.81 & 145.00 & 135.70 & - & 93.98 & 4.89 &  - & 4.92 & - & 4.61 \\ 
 & Co & 27 & 5352.04 & 70.20 &  61.50 & - & - & 4.57 &  4.47 & 4.45 & - & 4.36 \\ 
\hline
61CygA & Ca & 20 & 5260.38 & 83.50 & 69.00 & - & 88.05 & 6.05  & 5.73 & 5.75 & - & 5.99 \\ 
 & Cr & 24 & 5238.96 & 44.30 & 32.20 & - & 32.00 & 5.16  & - & 4.89 & - & 4.88 \\ 
 & Mn & 25 & 6021.81 & 148.10 &  132.00 & - & - & 5.33  & - & 5.00 & - & - \\ 
 & Co & 27 & 5352.04 & 25.50 & 15.60 & - & - & 4.61  & - & 4.31 & - & 3.03 \\ 
\hline
HD22879 & Ca & 20 & 5260.38 & 12.50 &  9.90 & - & 12.53 & 5.85  & - & 5.65 & - & 5.76 \\ 
 & Cr & 24 & 5238.96 & - &  - & - & 2.72 & 4.96 & - & - & - & 4.68 \\ 
 & Mn & 25 & 6021.81 & 24.80 &24.40 & - & - & 4.40  & - & 4.29 & - & - \\ 
 & Co & 27 & 5352.04 & 5.40 &  - & - & - & 4.23  & - & - & - & 4.09 \\ 
\hline
\end{tabular}
\label{tab:lines_P4}
\end{table*}

\subsection{Lines}

Four lines, of \ion{Ca}{I}, \ion{Cr}{I}, \ion{Co}{I} and \ion{Mn}{I}, were chosen to make the tests. Out of the many lines used in Paper~IV for determining the abundances of these elements, we searched for those that were flagged with ``Y'' in Tab.~C36 in the \fgdwarf, \kdwarf, \fgkgiant\ and \metalpoor\ groups. The ``Y'' flag  means that at least 50\% of the stars in a given group used that line for the final abundance of that element.  The atomic data were taken from the fifth version of the line list created for the Gaia-ESO survey \citep{Heiter2015b}.  All have $\log gf$ values which have been evaluated to be of good quality \citep[for details see][]{Heiter2015b}. Their atomic data are listed in Table~\ref{tab:linelist}. The two elements,  \ion{Co}{I} and \ion{Mn}{I}, both have hyperfine structure (HFS). Those components are also listed.

Their profiles are plotted for each star in Fig.~\ref{fig:profiles}. One can see that the lines are mostly clean, and of weak to moderate strength. Note that the Cr line in the metal-poor star HD22789 is very weak and could not be measured by all methods. Unfortunately, due to the intrinsic difference of the four stars, this was the only Cr line available that could be analysed by at least one method in all four stars. 
The abundances derived in Paper~IV for each element, line and star studied in the present work  are indicated in Table~\ref{tab:lines_P4}.  In addition, we consider a reference value for comparison which is indicated in Tab.~\ref{tab:abund_ref}. This value corresponds to the average of the abundances obtained by all methods and is taken from the third column of Tables C2, C8, C20 and C33 of Paper~IV for 61CygA, Arcturus, HD22879 and the Sun, respectively. It is important to state here that during this work this value is used solely as a zero point. It is beyond the purpose of this work to reproduce that result or provide another updated reference value.

\subsection{Atmospheric models}
The atmospheric models are those employed by the analysis of the spectra in GES, as well as in Paper~III and Paper~IV. These are the MARCS models \citep{Gustafsson2008}, which are computed under the 1D-LTE assumption and assume the standard composition for $\alpha$-enhancement with respect to iron abundance.

\begin{table}[t]
\caption{Reference absolute abundances of the four chosen lines for each GBS, taken from Paper\,IV.}
\begin{tabular}{ccccc}
\hline\hline
 & A(Ca) & A(Cr) & A(Mn) & A(Co) \\ 
\hline
Sun & 6.28 & 5.51 & 5.34 & 4.83 \\ 
Arcturus & 5.82 & 4.97 & 4.41 & 4.45 \\ 
61CygA & 5.87 & 5.11 & 4.83 & 4.10 \\ 
HD22879 & 5.75 & 4.77 & 4.08 & 4.11 \\ 
\hline
\end{tabular}
\label{tab:abund_ref}
\end{table}

\section{Methods}\label{methods}
We considered six methods in total, all being previously used in Paper~III and IV. Three methods are based on synthesis and three on equivalent widths (EWs), which are summarised in Tab.~\ref{tab:methods} and  shortly described below.  For more details we refer to Paper~III and IV, as well as \cite{2014A&A...570A.122S} for the GES analysis of the UVES spectra.

\begin{table}[t]
 \caption{Summary of spectroscopic methods employed in this work. The name of the method, the approach (EW: equivalent width, synth: synthesis), the radiative transfer code employed and the wrapper code that uses the radiative transfer code (if applicable) are indicated. }
 \label{tab:methods}
{\small 
\begin{tabular}{c | c c c}
\hline 
name & approach  & radiative transfer code & wrapper \\
\hline 
iSpec & synth &  SPECTRUM & iSpec\\
ULB & synth & Turbospectrum & BACCHUS\\
SME & synth &  sme\_synth & SME\\ 
Porto & EW & MOOG & \\
BOL& EW & SYNTHE & GALA\\
UCM & EW & MOOG & StePar\\
\hline
 \end{tabular}
 }
 \end{table}

\subsection{Synthesis Methods}
\subsubsection{Universit\'e Libre Bruxelles (ULB)}
The results of this method have been provided by T. Masseron and are based on BACCHUS \citep{2016ascl.soft05004M}, which was developed in Brussels. In this work, the method was based on finding the  synthetic spectrum from Turbospectrum \citep{2012ascl.soft05004P} that reproduces  best the observed spectrum with  $\chi^2$ minimisation between synthesis and observation. Each line is treated independently, and therefore a pseudo-continuum is used to normalise locally with a linear function the spectral region  of 20~\AA\ around that given line.   The method also is able to { automatically }determine the mask (spectral regions) that will be used for { calculating } the $\chi^2$. 

\subsubsection{Spectroscopy Made Easy (SME)}
\sme\ was developed in Uppsala and in the USA \citep{1996A&AS..118..595V, 2016arXiv160606073P} and has been used by S. Buder for this work.  \sme\ is written in IDL and performs a $\chi^2$ { minimisation} using synthetic spectra computed on-the-fly. The radiative transfer code is an integral part of the  \sme\ package. The fitting is done in pre-defined regions which can be a spectral segment, an absorption line or simply a region of the line. The synthesis is done locally around that line, in a segment that is pre-defined by the user. The observed spectrum is normalised locally for that given segment, taking pre-defined continuum regions and fitting a straight line to the selected high flux points in these regions.  

\subsubsection{interactive Spectroscopic Framework (iSpec)}
\ispec\ was developed by \cite{2014A&A...569A.111B} and was used by K. Hawkins for this work. {\tt iSpec} follows the procedure of \sme\ but it is written in python and calls the radiative transfer code SPECTRUM \citep{1994AJ....107..742G} in this case.  The continuum is also determined locally, but fitting to the pre-defined continuum regions a function that can be either a spline or a polynome of an order specified by the user. Here a second order polynomial was used for the local normalisation.

\subsection{Equivalent Widths}
\subsubsection{Bologna (BOL)}

This method has been developed at the University of Bologna and the results obtained with it in this work were provided by L. Casamiquela.
In this method, two steps are performed with two different codes. First, the EWs themselves are measured using DOOp \citep{2014A&A...562A..10C} which is an automatic wrapper for DAOSPEC \citep{Stetson2008}. The second step is to input the EWs into GALA \citep{2013ApJ...766...78M}, which uses the radiative transfer code SYNTHE \citep{2005MSAIS...8...76K} and optimises the physical parameters using the classical spectroscopic methods based on iron lines. Afterwards it measures the  line abundances.  Continuum subtraction in this case is performed for the determination of EWs globally, as described in \cite{Stetson2008}.
 
\subsubsection{Universidad Computense de Madrid (UCM)}

This method has been  used by H. Tabernero for this work. As for \bol, two different codes are considered: EWs are determined with TAME \citep{2012MNRAS.425.3162K} and the abundances with MOOG  \citep{sneden} { version 2014}. The continuum is also normalised globally for the EW determination. TAME does not allow to determine EWs without normalising the spectra, i.e. TAME always performs a re-normalisation of spectra for the EW determination.  The code {\tt StePar} \citep{2012A&A...547A..13T} is employed as a wrapper code to call MOOG and derive abundances in an automatic way.

\subsubsection{University of Porto (Porto)}
This method has been used by V. Adibekyan for this work and is  based on ARES v2 code \citep{sousa_ares} for the EW determination and on 
the 2014 version of the MOOG  radiative transfer code for the abundances.
The measurements of the EW is done locally around each spectral line, by performing a local normalisation of spectra. As in case of TAME (above), 
ARES also does not allow EW determination without re-normalisation of the spectra. For details on derivation of stellar parameters and chemical abundances
with ARES+MOOG we refer the reader to \citet{Sousa-14, Adibekyan-12}.

\section{Results}\label{results}
As in Papers\,III and IV,  our results were obtained by fixing the stellar parameters ($T_{\textrm{eff}}$ , $\log g$, [Fe/H], \vmic, \vsini) to the recommended values (see Table~\ref{t:bs_ap}). { We performed eight different tests listed below.}
\begin{enumerate}
\item {\it Linelist}: To understand the effect on final abundances that different methods have in dealing with a shifted wavelength of an element in the linelist.
\item {\it Continuum normalisation}: To study how the different methods normalise their data
\item {\it Hyperfine structure}: To quantify this, often neglected, quantum effect on final abundances. 
\item { \it Microturbulence}: To study the dependence of the final abundances on this parameter.
\item {\it $\alpha$-element abundances and continuum opacities}: To study how the different methods treat the $\alpha-$enhancement in the spectral modelling. 
\item {\it Atmospheric model interpolation}: To test the impact on abundances when using different interpolated atmosphere models. 
\item {\it Blends}: To study the impact of subtle blends on the final abundances by different methods.
\item {\it Radiative transfer codes}: To compare the radiative transfer codes using otherwise same data. 
\end{enumerate}
We aim at making a systematic study of these effects, which are commonly not discussed in the literature because they are believed to have a small impact in the final abundances. To do so, we determined abundances for each of these tests letting the rest of input parameters fixed. { We comment here that they are by no means a complete list of possible tests to investigate effects of abundance determination using different methods. There are many other effects which are not explored in this work but should be explored in the future in a similar way than here. One example is the effect of using self-consistent model atmospheres when abundances of some elements are more extreme, such as second generation globular cluster stars. Other examples are the use of different source data for the line modelling, such as oscillator strengths, line opacities in the source function, the partition functions or the damping approximations. For the latter, in our case we used the damping constants provided in the GES linelist, that is, van der Waals constants of \cite{BPM} for Ca, Cr and Co, and the semi-empirical calculations of \cite{K07} for Mn.  In this work we have kept this information fixed but further effects of abundances in this direction are interesting to explore. }

{ The result for the 8 tests performed here } are discussed separately in the following sections.

\subsection{Linelist}
\label{1.linelist}

\begin{figure}
\hspace{-0.5cm}
\includegraphics[scale=0.6]{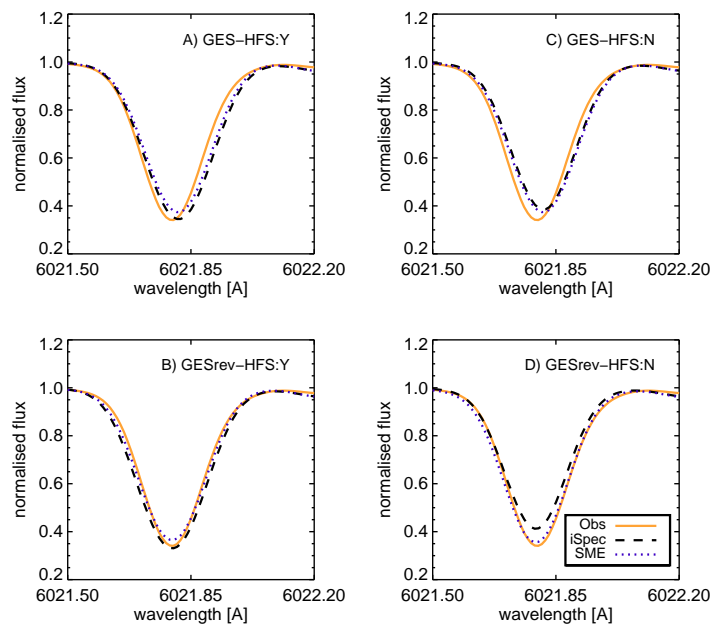}
\caption{Profiles of Arcturus of Mn line. Each panel represents the syntheses of \sme\ and \ispec\ corresponding to the final abundance in each of the runs. }
\label{fig:mn_profiles}
\end{figure}

We found that the wavelength of the selected Mn feature was incorrect in the GES linelist version 5.0. This was realised because no extra local RV correction was applied to the synthesis methods and therefore the core of the Mn line in the synthesis of each star was slightly shifted (by the same amount) with respect to the observation.  An example is shown for Arcturus in the upper panels of \fig{fig:mn_profiles}, in which we show the syntheses of \ispec\ and \sme\ with and without hyperfine structure splitting (HFS) along with the observed profiles of Arcturus.
To build the new hyperfine structure list we considered the weighted mean position of the line (hence the weighted median upper and lower energy levels of the transition; the HFS constants A and B of each energy level; the angular momentum quantum number of the levels (J) and the nuclear spin of the atom (I), and the total $\log gf$). 
Applying the Casimir formula \citep{casimir1936interaction, schwartz1955theory} { and the Russell-Saunders formulae \citep{condon1935theory} } for relative line strength allows us to derive each component and its respective $\log gf$. Although the HFS constants and $\log gf$ values are those from the GES linelist, the HFS components in the GES linelist around the Mn line are binned to 10~m\AA, {  while here no binning was applied}. This allows us to test not only the effect of the overall line shift, but also the apparent {\it number} of components of the HFS.  The lower panels of \fig{fig:mn_profiles} show the syntheses of \sme\ and \ispec\ of Arcturus with the revised line data along with the observations.   One can see that the core of the line agrees in the new case. 

\begin{figure}
\hspace{-0.5cm}
\includegraphics[scale=0.65]{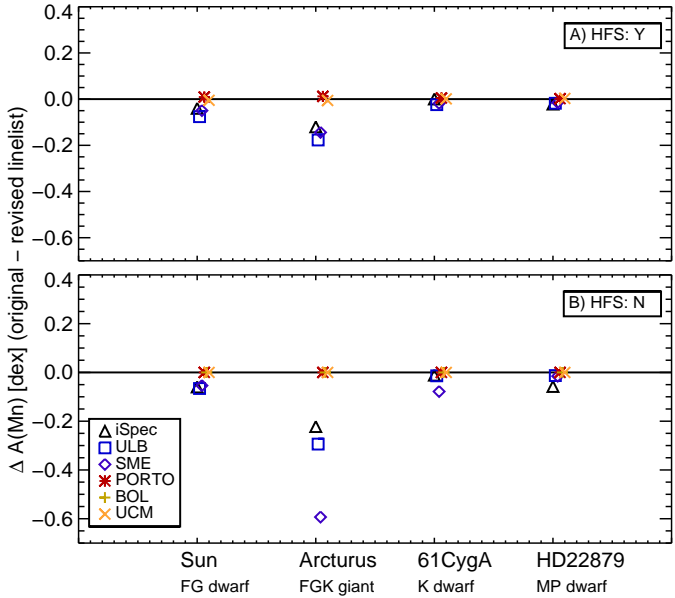}
\vspace{-1.5cm}
\caption{Difference in Mn abundances per star per method between the original Mn wavelength and the correct Mn wavelength for a) considering HFS of Mn; and b) not considering HFS.}
\label{fig:mn_offset}
\end{figure}
 
In automatic determination of elemental abundances of large high resolution spectroscopic surveys, it is highly probable that some atomic transitions will have laboratory wavelengths that do not perfectly agree with the observations. Furthermore, it is known that atmospheric motions can cause slight shifts of the position of atomic lines, also producing a slight disagreement between the laboratory wavelength and the observation \citep{2012A&A...544A.125M}.  Finally, sometimes wavelength calibration problems exist, shifting some lines from the laboratory \citep[see also extensive discussion on the impact of determining stellar parameters and abundances when issues of wavelength calibration arise in][]{2016arXiv160703130H}. Since many methods perform only a global RV correction, abundances determined from shifted lines can be more uncertain. In this section we quantify this effect. 

Furthermore, since Mn is an element that is affected by HFS, the splitting in the line list can be binned in different ways. In addition to the different wavelength position of the core of the line, the GES linelist is { included} 6 components,  { while the corrected line included 15 components. This allows us to}  study how the number of components considered for the HFS splitting affect the final abundance.

The abundances of Mn were derived four times for each star by each method as follows:
\begin{enumerate}
  \item No HFS, linelist version 5.0 (incorrect wavelength),
  \item No HFS, GES linelist version { 6} (correct wavelength),
  \item With HFS, GES linelist version 5.0  (incorrect wavelength),
  \item With HFS, GES linelist version { 6} (correct wavelength).%
\end{enumerate}
The results of these 4 runs are shown in Fig.~\ref{fig:mn_offset}, in which the abundances determined from the old and the new linelist were compared for each star and method. In the upper panel we compare the results when HFS is taken into account (hereafter HFS:Y), while in the lower panel we compare the results when HFS is not taken into account (hereafter HFS:N).

In agreement with the results obtained in \cite{2016arXiv160703130H}, we found that when considering HFS:N, we can see in the bottom panel of \fig{fig:mn_offset} that the EW methods (\porto, \ucm\ and \bol) do not show a difference in the derived abundance when the linelist presents a small shift in wavelength. This is not surprising, as  abundances determined from EW consider the curve of growth, which relates the strengh of a line with the abundance. Thus,  a small shift in the line list does not affect the measurement of the line strength, i.e. the EW itself.   For the spectrum synthesis methods (\sme, \ispec\ and \ulb), on the other hand, a difference in the derived abundance is seen.  This is also expected since synthesis methods { perform a pixel-by-pixel fitting between model and observation} and therefore it is important that both profiles are properly aligned. The magnitude of the difference varies between the stars and methods, where weak lines like the metal-poor star show small differences of the order of 0.06~dex by \ispec\ and 0.01~dex by \sme\ and \ulb, the stronger lines like the one of the giant can produce significant differences of up to 0.6~dex by \sme. Note that even in the case of the Sun, a difference of 0.06~dex by \ulb\ in retrieved abundances can arise for a bad position in a given line.  In the right hand panels of \fig{fig:mn_profiles} one can see how the fit improves from Panel C to D, in particular for \sme, when the line is properly aligned for synthesis and observation.

When considering \hfsy, we can see in Panel A of \fig{fig:mn_offset} that even the EW methods are { slightly} affected. Although the same EW is used in all cases, the different components in the HFS produce an  abundance { that differs by up to 0.01~}dex in the case of Arcturus when using the \porto\ method. We note here that \bol\ results are not shown, because this method cannot take into account HFS.  For the synthesis methods the differences are less significant than when we consider \hfsn, which is due to a smaller overall shift of model and observation.  Interestingly, the extreme difference of 0.6~dex shown by \sme\ in the \hfsn\ case drops to less than { 0.1}~dex for the \hfsy\ case.  This can be seen in the left hand panels of \fig{fig:mn_profiles}, in which the left wing of the line is less affected by this shift when the synthesis takes HFS into account. 

It is instructive to comment that this issue is well-known and therefore most synthesis methods have a way to deal with it. For example, \sme\ has a functionality implemented which corrects locally the radial velocity of the star, in order to align the synthetic with the observed profile. Other methods without this implementation, such as \ulb\ and \ispec\ in this work, but also  several other ones reported in the literature, especially those based on a grid of synthetic spectra \citep[e.g.][]{2006MNRAS.370..141R, 2015arXiv151007635G}, require a more detailed preparation of the line list. 

A common approach is to generate ``astrophysical calibrations" to correct oscillator strengths, damping coefficients, and in several cases the { central wavelength} of the line. Such calibrations are done with fitting synthetic spectra to spectra of well-known stars (the Sun and Arcturus for instance) but determining the aforementioned line list parameters \citep[see e.g.][]{2015ApJS..221...24S, 2016A&A...587A...2B}. This ensures that the line will provide accurate synthetic spectra for the analysis. Other option is to reject bad lines based on the confidence of the atomic data from the laboratory measurements, as well as the good reproducibility of the line profile of the synthesis of these well-known stars \citep{Heiter2015b}. This is the procedure implemented for instance by GES. { In both approaches a fraction of lines with inaccurate parameters may remain after processing, due the large number of lines that are observed by current high-resolution spectrographs.} We realised here that even in our highly detailed analysis of GBS based on the careful preparation of the GES linelist, the Mn line had a core wavelength that was shifted. Such issues are best assessed with an extra correction, as that one made by \sme. This is however difficult for grid-based methods, such as  MATISSE \citep{2006MNRAS.370..141R} or the ASPCAP pipeline of APOGEE \citep{2015arXiv151007635G}.

\begin{figure*}
\includegraphics[scale=0.9]{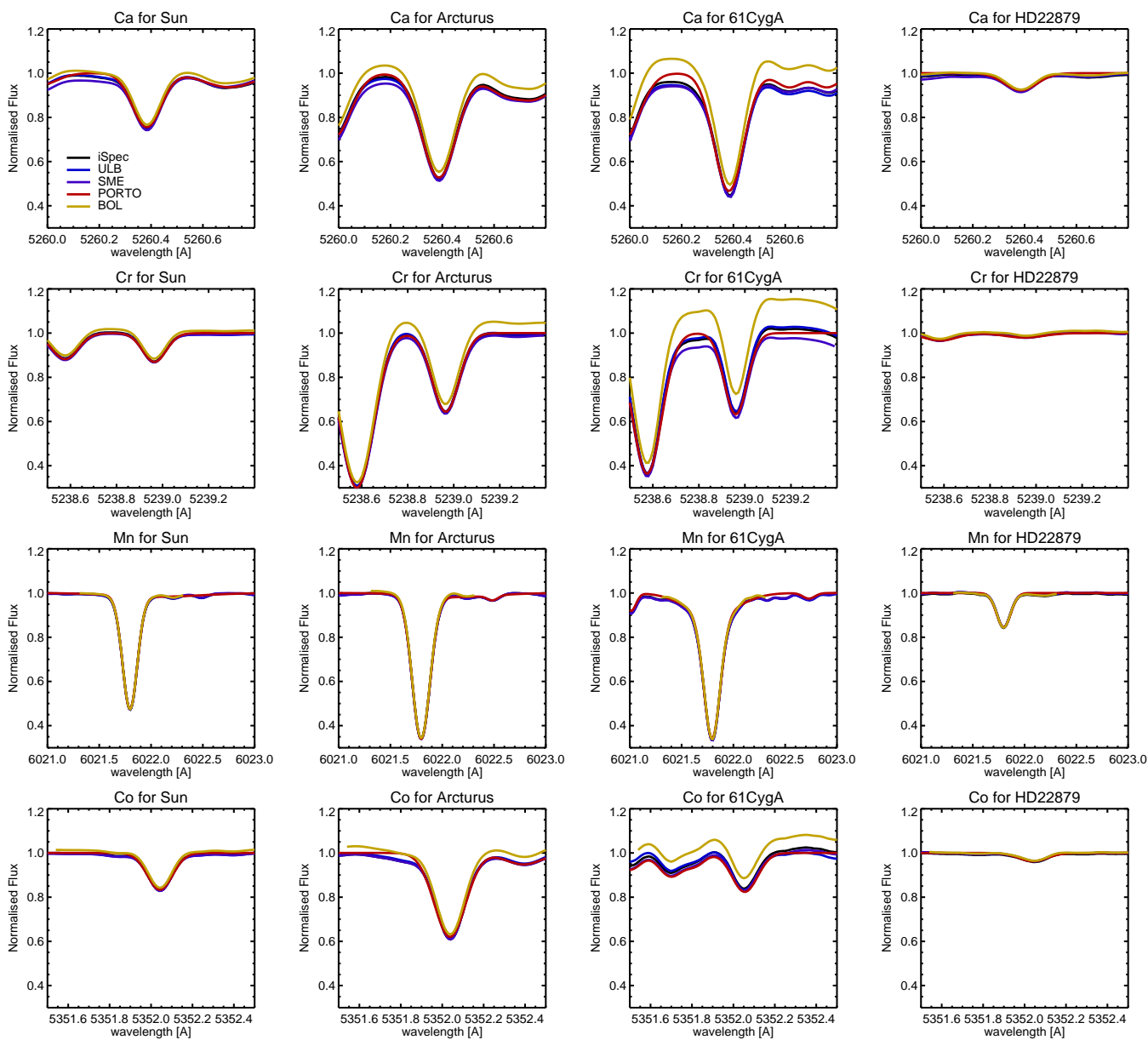}
\caption{Spectra of all GBS and lines used in this work as provided after the global normalisation performed with \ispec. Re-normalised observed spectra are shown with different colours for the different methods, as indicated in the legend. }
\label{fig:cn_normprof}
\end{figure*}

\subsubsection*{Conclusion for line list test}
EW methods are robust for shifts in wavelength in the linelist, but synthesis methods can be significantly affected and easily produce differences of 0.2~dex. To overcome this issue, synthesis methods should always perform an extra local wavelength correction to ensure proper alignment of model and observation, on top of using a well-tested line list.  
{ In addition, binning of HFS components in wavelength may have an effect on abundances, although it seems to be small in the case of EW methods. Further studies for a wider variety of elements and transitions are needed to assess this issue}. Hereafter the tests will always be made with the revised Mn linelist.

\subsection{Continuum normalisation}
\label{2.Cont_norm}

The vast majority of the spectral analyses methods to determine atmospheric parameters and chemical abundances need to perform a normalisation of the continuum flux. This fundamental step is done in several different ways, such as by fitting global polynomials to the pseudo-continuum \citep{2006ApJ...636..804A,  Stetson2008, 2010A&A...517A..57J, 2015arXiv151007635G}, to more local functions using synthetic spectra or linear fits, such as the syntheses methods employed in this work. Other methods do not normalise the continuum and perform the parameter determination adding an extra parameter to fit for this function at each pixel \citep{1998A&A...338..151K, 2009A&A...501.1269K}. Since there is still no consensus on the best way to treat the continuum flux of the data, it is well known that the measured abundances carry an error due to normalisation.  However,  since this process is done in such a large variety of ways, it is not always possible to quantify this source of error in determination of elemental abundances. 

\begin{figure*}[!t]
\includegraphics[scale=0.6]{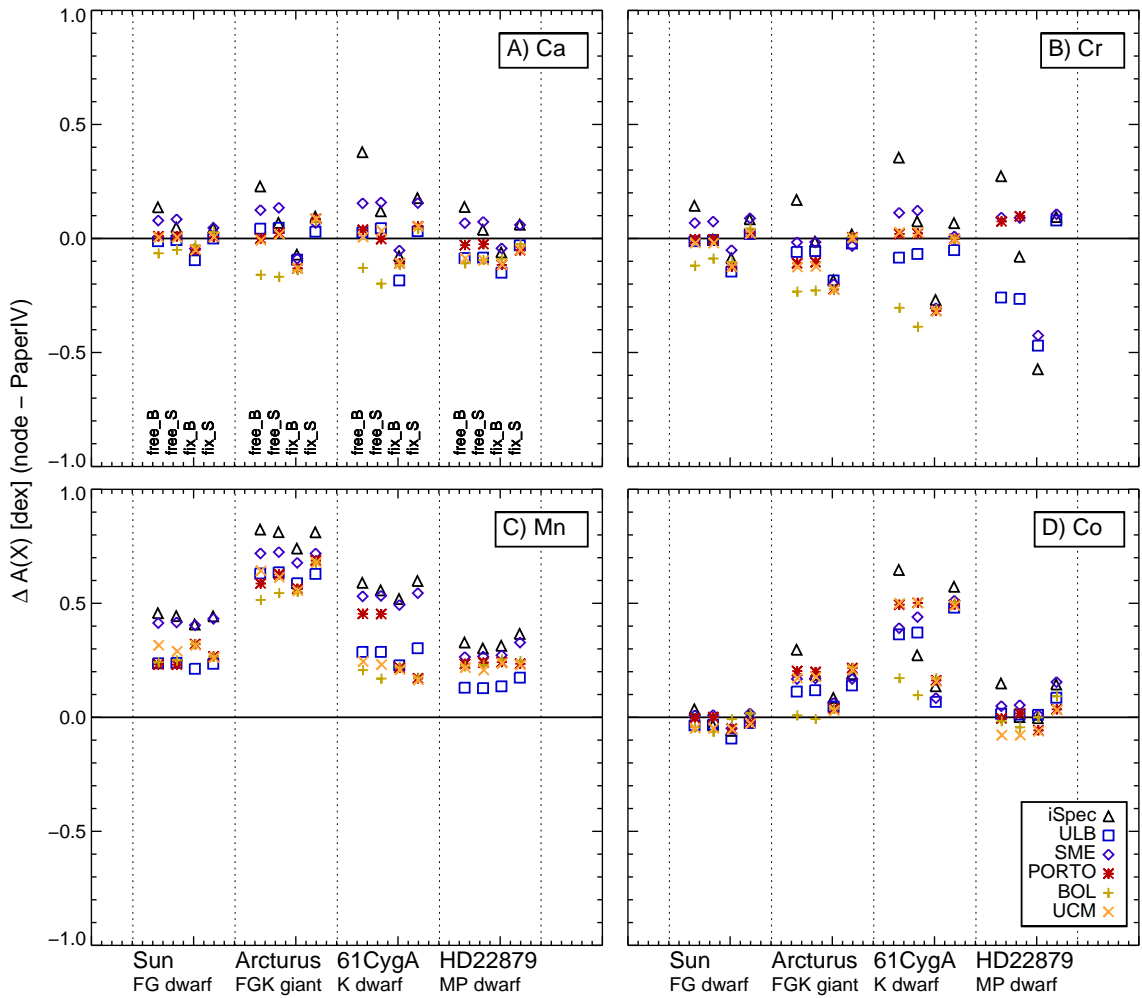}
\vspace{-1cm}
\caption{Difference in element abundance derived by each method for each normalisation test compared to the abundances of \tab{tab:abund_ref} for each element in a different panel. }

\label{fig:cn_abundcomp}
\end{figure*}

In this test we attempt with our methods to estimate typical uncertainties due to continuum normalisation. To do so, the abundances of each element were determined four times for two different normalisation procedures ($S$ and $B$, see below) as follows: 
\begin{enumerate}
  \item each method performed own normalisation to the provided spectrum $S$ ($free\_S$),
  \item each method analysed a fixed spectrum $S$ and performed no further normalisation ($fix\_S$),
  \item each method performed own normalisation to the provided spectrum $B$ ($free\_B$),
  \item each method analysed a fixed spectrum $B$ and performed no further normalisation ($fix\_B$).
\end{enumerate}
Here the abundances of Mn and Co were determined without considering HFS. This allowed a comparison of all methods that only concentrated on the continuum normalisation issue.  We further remark here that the spectra used in this work have already been pre-normalised as described in Paper~II, i.e. using the synthetic spectra with stellar parameters of the star to find the continuum points and then fitting cubic splines to the local pseudo-continuum. Thus, any further normalisation test essentially meant performing a re-normalisation of the provided spectrum.  In this sense, when we refer to the $S$ spectrum we consider the original spectrum as normalised by {\tt iSpec}, while the $B$ spectrum refers to the spectrum of {\tt iSpec} re-normalised by {\tt BOL}.   We mention that \bol\ renormalised the entire continuum, including the Cr line in HD22879, but did not measure abundances because the line is too weak.  We also mention that in the $free$ test, the methods \ispec\ and \bol\  performed a re-normalisation of their already normalised spectrum. In the case of the $S$ spectrum, the continuum was normalised globally using a synthetic spectrum with the stellar parameters of the star (see Paper~II), which could be further normalised locally around the line into study. In the case of \bol, an iterative process in measuring EWs and fitting the pseudo-continuum is performed \citep{Stetson2008}, which allows a re-normalisation of the spectrum $B$ by \bol\ in this particular test.

We further comment that re-normalising the pseudo-continuum is not unusual in the field. Several GES nodes perform a re-normalisation onto the data products \citep[see ][for node despcription]{2014A&A...570A.122S}. Indeed, some methods such as ARES (\porto) or TAME (\ucm) are not able to determine EWs without normalising and therefore, whenever they analyse a normalised spectrum (like in this case) they re-normalise. 
To overcome this issue in this and the following tests using fixed continuum, \porto\ and \ucm\ determined the abundances considering the EWs measured by \bol, which is able to determine EWs from fixed normalised spectra. 
 
Figure~\ref{fig:cn_normprof} compares the freely re-normalised observed spectral profiles produced by each method. The 16 panels show each element per star.  For the Sun and the metal-poor star HD22879 the normalised spectrum produced by each method is almost identical to the input spectrum. For the Ca feature for both stars there is some slight variation in the placement of the continuum to the left of the spectral line.

For Arcturus and the K dwarf 61CygA there are more gross differences in the normalisation. \bol\ in particular consistently places the continuum lower than the other methods for these stars for the Ca, Cr and Co spectral features. This is due to the effective continuum placement technique \citep[see][Sect. 3.2]{Stetson2008} that places a depressed continuum  which tries to take into account, in a statistical way, the unrecognised flux deficits and excesses due to contamination effects. For Mn \bol\ provides a continuum placement matching those of the other methods and a good fit to the input observed spectrum. The normalisation of the spectral features in 61CygA shows the most difference between the methods. \porto\ underestimates the continuum placement for the Ca feature, while \sme\ overestimates the continuum placement for the Cr feature. 

One can see from all panels how the global fit performed by \bol\ is very different from the rest, with a systematic underestimation of the continuum with respect to the rest of the methods. This is the reason why we decided to consider this normalised spectrum for comparing $B$ and $S$.

Figure~\ref{fig:cn_abundcomp} shows the abundances for each star derived in this test differenced to the reference element abundances listed in \tab{tab:abund_ref}. Each panel shows one element and each star is separated by the dashed vertical lines. The four tests are shown next to each other, following the same order indicated in Panel A.  

When fixing the continuum (2 right hand tests for each star/element) we can see that the node-to-node scatter in the abundances decreases. Furthermore, this scatter does not seem to depend heavily on the input normalised spectrum, which remains very similar for the $fix\_B$ and $fix\_S$ cases. The zero point, however, significantly changes, with the results obtained from the $fix\_B$  spectrum, being systematically lower than those obtained from the $fix\_S$ spectrum. This can be explained by looking at the profiles of both spectra, where the continuum of the $B$ spectrum is systematically higher than the $S$ spectrum (see \fig{fig:cn_normprof}).  

In the case of each method performing own normalisation to input spectra, we see in the left hand tests for each star/element in \fig{fig:cn_abundcomp} that the node-to-node scatter is higher when the $B$ spectrum is used with respect to the $S$ spectrum. This suggests that using the spectrum which requires larger continuum corrections by each method (the $B$ spectrum) results in a larger range of abundances.

 It is interesting to note that when comparing the $free$ and $fix$ tests using $B$ spectrum, the node-to-node scatter decreases significantly, but also the mean absolute value changes. The latter is because when the nodes apply their normalisation for $fix\_B$ the fit to the observations improves and the abundances are closer to that of the method which provided the observations (\bol). While the node-to-node scatter of the $S$ spectrum from the $free$ to the $fix$ test improves, the zero point does not change significantly. Because the $S$ spectrum is already normalised well to the continua provided in Paper~IV, adding a correction to the normalisation in this case introduces only noise in the measurement.  The mean $fix\_S$ is close to \ispec, analog to the finding that the mean $fix\_B$ is close to \bol.
 
 Another interesting feature to note from \fig{fig:cn_abundcomp} is that for  Arcturus and \cygA, a large node-to-node scatter for all determination of abundances is obtained in the $free$ test, reflecting the well-known challenge of normalising cool stars, which are very crowded by absorption lines \citep[e.g. ][as well as Paper III and IV for extensive discussions]{2012A&A...547A.108L}. On the contrary, HD22879 has a smaller node-to-node scatter, since it presents few and weak lines, facilitating continuum normalisation. Still, freeing the continuum can  produce larger node-to-node scatter for very weak lines, such as the Cr one (see \fig{fig:cn_normprof}) than when the continuum is fixed. 
 
\begin{figure}[!t]
\includegraphics[scale=0.6]{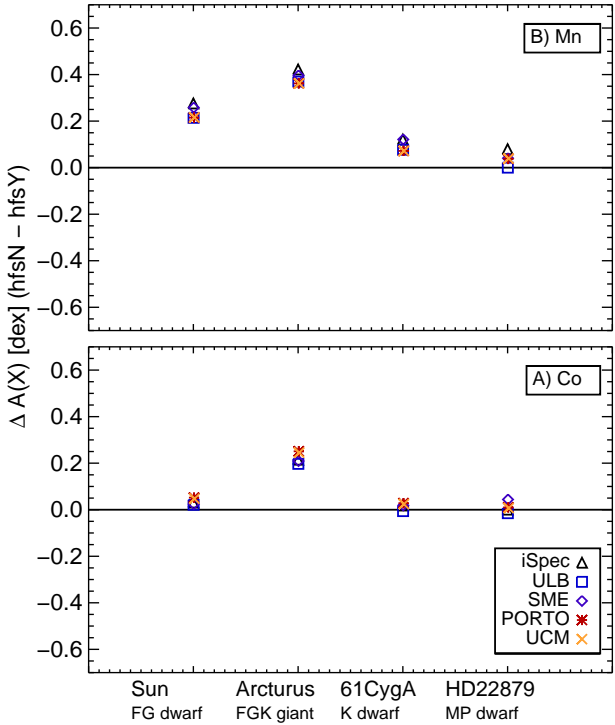}
\vspace{-1cm}
\caption{Difference in Mn and Co abundance derived by each method with and without HFS.}

\label{fig:hfs_abundcomp}
\end{figure}

 Finally, we comment here the significant differences of the abundances determined in this work with respect to those of \tab{tab:abund_ref}, such as Mn for all stars and Co for \cygA. One difference is  that here HFS was neglected, while the reference value is only from methods with HFS. See Fig. 11 in Paper IV showing the size of the effect for the Sun, which is similar to that seen here.   See also \sect{hfs} and \fig{fig:hfs_abundcomp}.  Another difference is that here we employ the revised line list while in Paper~IV the old line list was employed, and the abundance of Mn was provided by \ulb\ among other methods. As discussed in \sect{1.linelist},
 \ulb\ does not perform a further wavelength correction to the line. We recall  that the final abundances for Mn of GBS { are based on several Mn} lines but a revision of this line should be done in future releases of abundances of GBS. In this context, we also remind the reader  that we do not intend to correct nor reproduce the reference abundances of Paper~IV in this work, here we only use them as a comparative value for assessing the node-to-node scatter.  The final Mn abundances for the GBS in Paper~IV follow a different strategy of performing differential abundance with respect to the Sun, aiming at reducing the systematic uncertainties such as continuum placement (see Paper~IV for extensive discussion).  The values used here are the average of the absolute abundances obtained by each method that considered HFS, and do not reflect the final { differential} abundances obtained in Paper~IV suitable for calibration purposes.

\subsubsection*{Conclusion for Continuum Normalisation test}
The agreement between methods is improved if the continuum of a spectrum is fixed, although the final absolute values highly depend on which continuum is applied to the data.  Therefore, a careful continuum normalisation should be performed  which then should be kept fixed for abundance determination.  Hereafter our tests will always be made with the  $fix\_S$ method, which means that abundances determined by \ucm\ and \porto\ will consider the EWs determined by \bol. 
 
\begin{figure}[!t]
\includegraphics[scale=0.6]{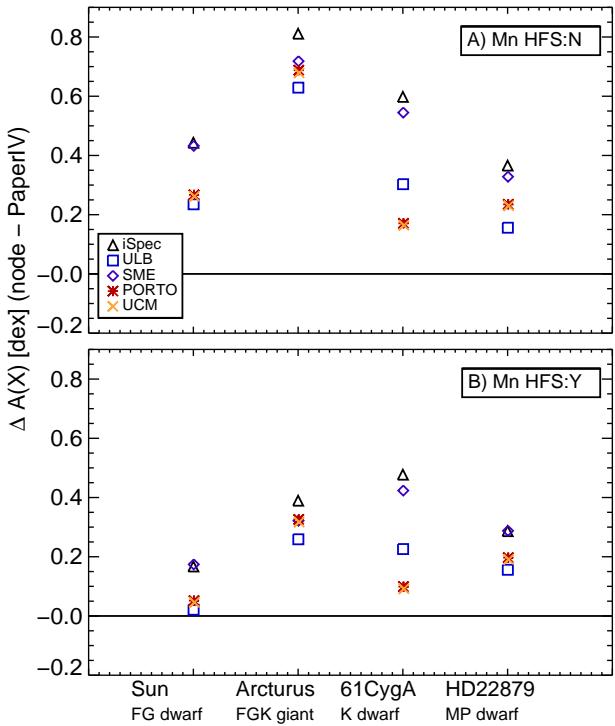}
\vspace{-1cm}
\caption{Difference in Mn with the reference value for each method with and without HFS. }

\label{fig:hfs_abundcompA}
\end{figure}

\subsection{Hyperfine structure splitting (HFS)}\label{hfs}

Current high-resolution spectrographs like UVES, APOGEE and GALAH  can  fully resolve many features in
absorption lines. On top of the basic atomic data of wavelength and oscillator strength, information on
line broadening  due to quantum effects such as HFS can be disentangled.
Since HFS causes the line profile to increase in width and the
peak intensity of the line to decrease, the line appears
asymmetric and can no longer be described by a simple Voigt
profile  \citep{2005ApJS..157..402B}. Neglecting this results in the
incorrect measurement of the wavelength and a miscalculation
of the equivalent width or synthetic spectrum, leading to an incorrect calculation of the abundance of this given element. 

Manganese is known to be strongly affected by HFS producing differences in retrieved abundances of up to 0.6~dex \cite[e.g.][and Paper~IV]{2005A&A...441.1149D, 2012A&A...541A..45N}.   Studies of HFS effects on other odd-Z elements such as Co also show that differences in derived abundances of Co can be up to 0.1~dex  \citep[][see also Paper~IV]{2005A&A...441.1149D, 2008ARep...52..630B}. Although the majority of modern studies of elemental abundances take this effect into account, there are  still some cases in which HFS is not implemented \citep[e.g.][to name a few]{2009A&A...497..563N, 2015A&A...574A..50J, 2015A&A...583A..94A}. One example in our work is the \bol\ method, which derives abundances from EWs with the radiative transfer code SYNTHE.  This code does not support calculations of HFS.  For this reason, the \bol\ method is not considered in the discussion of this section. 

Here we attempt with our methods to estimate typical uncertainties due to HFS. To do so, the abundances of Mn and Co were determined twice as follows: 
\begin{enumerate}
  \item each method determined abundances considering hyperfine structure splitting (HFS:Y)
   \item each method determined abundances of  neglecting hyperfine structure splitting  (HFS:N)
\end{enumerate}

In \fig{fig:hfs_abundcomp} we compare the abundances obtained by each method for each star for Mn and Co, where we plot the differences of the abundances obtained from both runs for Mn in the upper panel and Co in the lower panel. Each method is represented by a different symbol and colour as indicated in the legend, which follow the same symbols and colours as previous figures. We can see that for both cases, HFS:N systematically overestimates the abundance by a value that depends on the spectral type and element. Mn shows large differences of up to { 0.4~dex} in the case of Arcturus. The differences in abundances between HFS:N and HFS:Y { are similar for all methods}, with the metal-poor star presenting no differences  at all for \ulb. { The fact that the differences are significantly larger for Arcturus than for HD22879 indicates how the HFS effect depends on line shape: stronger, more asymmetric lines with broader wings result in  larger differences in abundances when the modelled line is split into its HFS components.} We remark here that even for the Sun a difference of { 0.2~dex} in Mn abundance can be obtained due to HFS. 

The HFS of Co is more subtle  than for Mn, and therefore for Co the differences are in general smaller than for Mn, below 0.05~dex for most of the cases. Arcturus, which is the star with the overall stronger lines (see \fig{fig:profiles})  is an exception presenting a difference of more than 0.2~dex.


Figure~\ref{fig:hfs_abundcompA} shows another striking result: while the abundances obtained under \hfsy\ are systematically lower than the results obtained under \hfsn, the node-to-node scatter does not seem to significantly decrease.   This also shows that even if HFS is taken into account, the abundances might be uncertain due to the HFS treatment, binning of the different components in the line list (see \sect{1.linelist}) and the proper modelling of lines that strongly deviate from having Gaussian or Voigt profiles. The line under study, even in the case of the Sun can produce 0.2~dex differences in abundances obtained by different methods.

 \begin{figure*}[!t]
\includegraphics[scale=0.4]{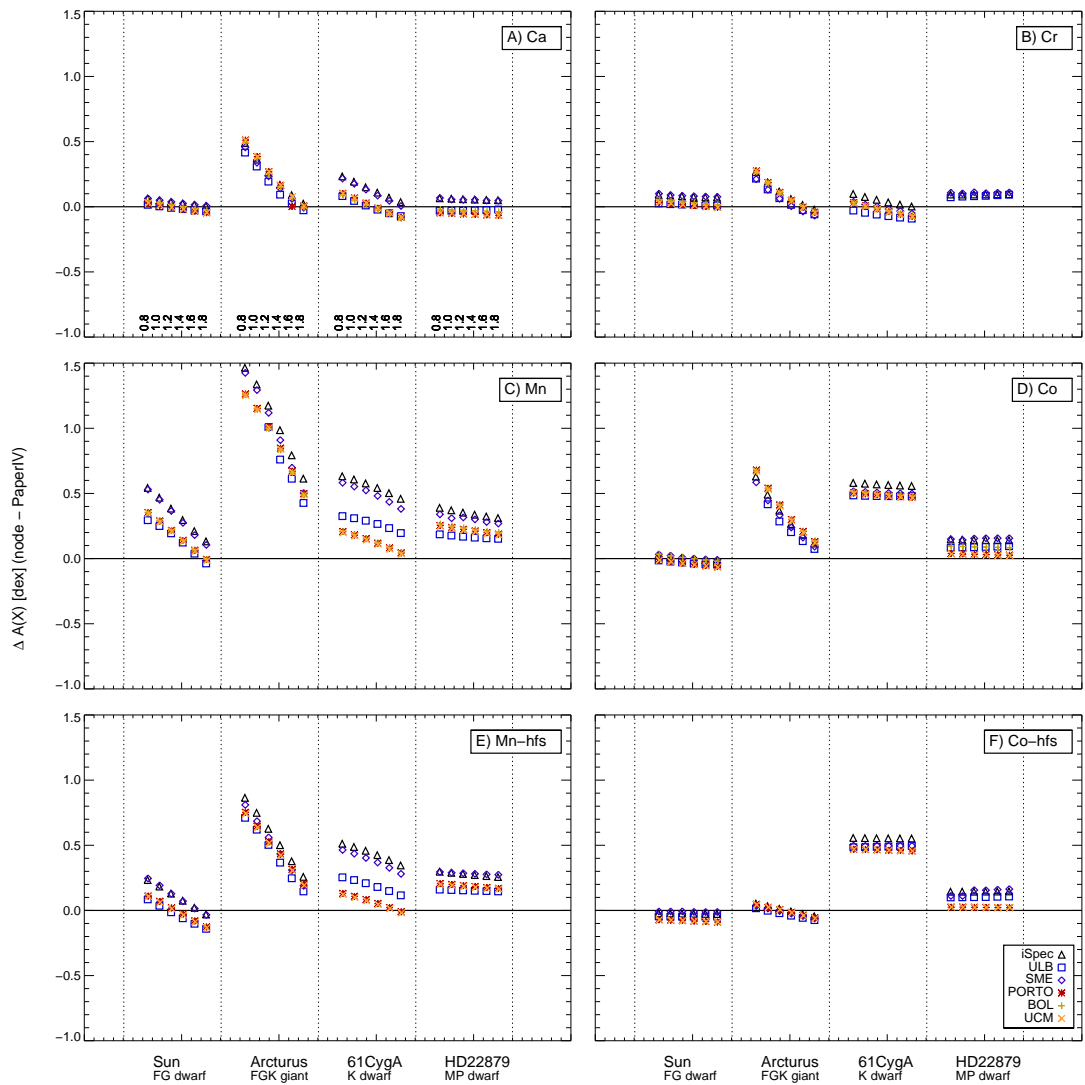}
\vspace{-2cm}
\caption{Difference in element abundance derived by each method for each value of microturbulence compared to the reference abundances for each element.}
\label{fig:vt_abundcomp}
\end{figure*}

\subsubsection*{Conclusion for HFS}

Overabundances of the order of { 0.4}~dex can appear in strong Mn lines  in giants if HFS is neglected, but even for weak lines, overabundances of 0.15~dex can be obtained. The HFS effect of Co is less significant but still for giants it can be of 0.2~dex. These differences are much larger than accuracies of 0.05~dex in iron-peak elements needed for disentangling different stellar populations, star formation histories and constraining chemical evolution models as well as nucleosynthesis yields \citep[e.g.][]{2007A&A...467..665F, 2013ARA&A..51..457N}.  
{ We would like to point out} that in the fourth release of GES (Hourihanne et al., in prep), HFS is not taken into account by all the nodes, adding in some cases uncertainties of the order discussed here. { The results of} this work have motivated the GES node \ucm\  \citep[see][]{2014A&A...570A.122S} to include HFS for future GES data releases, { which should improve the accuracy of the derived abundances of this survey.} 

\subsection{Microturbulence}

When deriving atmospheric parameters and chemical abundances using 1D model atmospheres, the microturbulence (\vmic) parameter also needs to be determined. { In 1D models, this parameter is meant to account for some of} for the turbulent motions in the atmosphere which cause the spectral lines to be broadened. A realistic description of this physical mechanism leading to an accurate modelling of the line profiles can only be done  in 3D. This is computationally expensive and currently only few of these accurate models are available to the community \citep{2013A&A...557A..26M}. The stronger the line, the more it is affected by this broadening and thus by \vmic. Because in spectral synthesis calculations under 1D \vmic\ is not physical, each code can employ slightly different values for \vmic\ given { an otherwise defined set} of stellar parameters (see Paper~III). { Moreover, it has been found that the microturbulence derived by minimising the trend of Fe line abundances against reduced equivalent width decreases with increasing metallicity  \citep[e.g. by][using APOGEE spectra of giant stars]{2016arXiv160408800H}} 

 In this section we attempt with our methods to estimate typical uncertainties due to different values derived { or assumed} for \vmic. To do so, the abundances of all elements were determined 6 times fixing  values of \vmic\ $ = [0.8, 1.0, 1.2, 1.4, 1.6, 1.8]~\mathrm{km/s}$.  The results for all stars, elements (with and without HFS for Mn and Co)  and methods are shown in  \fig{fig:vt_abundcomp}.  In the figure, Panel A shows the abundances of Ca for all methods and stars and, as in \fig{fig:cn_abundcomp}, the results of all runs are plotted side-by-side. The values of \vmic\ are indicated at the bottom of the panel and they follow the same order in each star and the rest of the Panels. The methods are indicated by different colours and symbols, which are explained in the legend and are the same as in previous figures. Panel B shows the results for Cr, Panel C and D show the results for Mn and Co neglecting HFS, and Panel E and F show the results for Mn and Co when HFS is considered.

The strong dependency of derived abundance on \vmic\  can be seen for  { almost } every star, element and method. { Exceptions are Cr and Ca for HD 22879 and Co for all stars except Arcturus. When there is a dependency},  it behaves in every case in the same way, in which the larger the \vmic\ value, the smaller the obtained abundance. This is because a large \vmic\ value will imply a stronger line and therefore less abundance is needed to model a line of a given strength.  { Note that for this test we varied the microturbulence by a large amount (1 km/s). For example, differences of 0.2~km/s as found in Paper~III correspond to differences of 0.1~dex in abundance for strong lines like those of Arcturus but of 0.05~dex or less for weaker lines.  }

{ All methods have a very similar dependency with respect to \vmic. }
In each panel we see that the strongest dependency of \vmic\ and abundances is found for Arcturus, because it presents the strongest lines of the stars in our sample. Likewise, the weakest dependency { of abundances on \vmic\ } is seen for HD~22879. For the weakest lines of HD~22879, like the Cr and Co line, the dependency of the derived abundance is almost null. Interestingly, a weaker dependency of \vmic\ with abundances is seen for Mn and Co when HFS is taken into account. This is due to the splitting into several components, \vmic\ plays a less important role as each composed acts as a weak line.  


\subsubsection* {Conclusion for microturbulence test}
There is a dependency of the abundances on the adopted value of \vmic, where larger abundances are derived for smaller values of \vmic. This relation is well known and several empirical \vmic\ relations exist in the literature  { \citep[e.g.][]{2005ApJS..159..141V, 2012A&A...547A..36A, 2013ApJ...764...78R}} which can be used when the spectra do not allow to determine this parameter together with the rest of the stellar parameters \citep[see discussions in e.g.][]{2014A&A...570A.122S, 2016arXiv160408800H}. As in the case of the continuum normalisation, the absolute value of the abundance { depends on the } adopted value of \vmic, but since this parameter is  not a physical parameter per se, it is difficult to find an absolute value for \vmic\ that would account for the correct line broadening in every method. We also found that if HFS is taken into account, strong lines become less dependent on \vmic.

  \begin{figure*}[!t]
\includegraphics[scale=0.6]{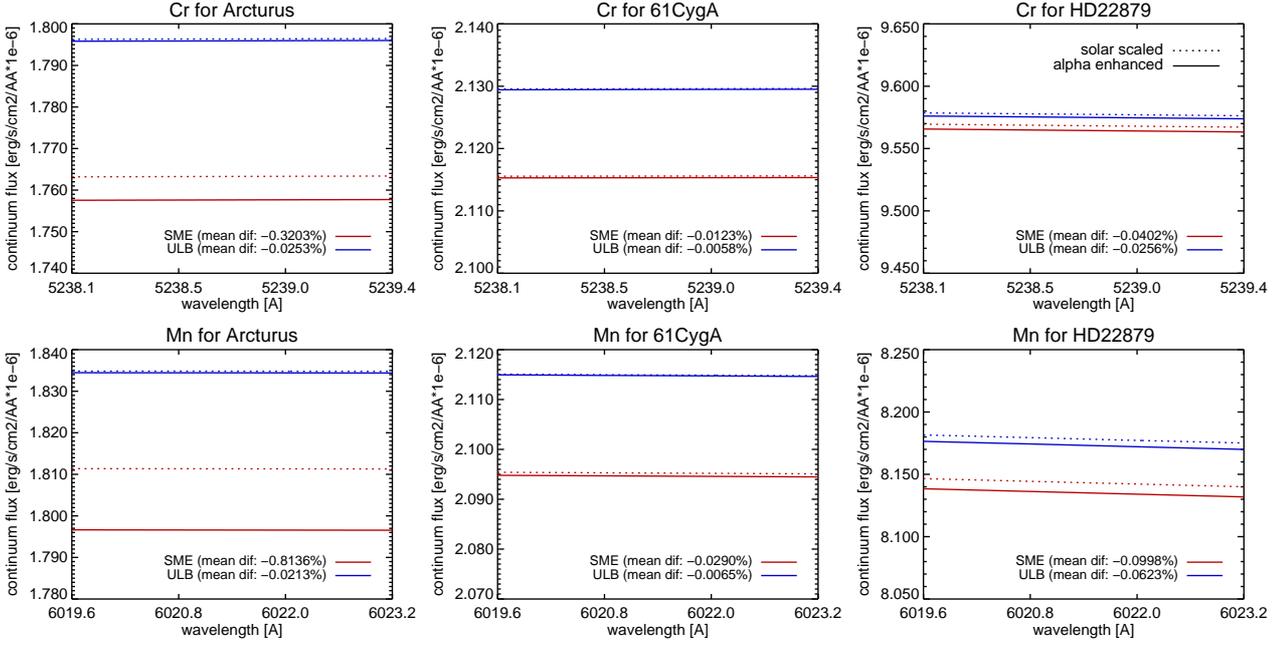}
\caption{ Syntheses performed with \sme\ and \ulb\ considering $\alpha-$enhanced and solar-scaled composition. The syntheses were performed without line absorption, and show how the continuum fluxes change when $\alpha-$enhancement is taken into account or not. The Sun has been excluded from this figure as  by definition it has zero $\alpha-$enhancement and the syntheses are identical. The bluest and the reddest line regions analysed in this work were chosen for this illustration. The difference of the fluxes (solar-scaled - $\alpha-$enhanced) in percentage are shown at the bottom of each panel. { The syntheses with $\alpha-$enhancement are plotted with continuous lines while the syntheses with solar-scaled abundances with dotted lines. The syntheses computed with \sme\ are plotted with red colour, while the syntheses computed with \ulb\ (Turbospectrum) are plotted with blue colour. } }
\label{fig:alpha_syn} 
\end{figure*}

\subsection{Enhancement of $\alpha$-elements and continuum opacities}

When solving the radiative transfer equations to derive chemical abundances of a star, there are three main inputs. The model atmosphere, providing the pressure and temperature of the gas at different depths in the atmosphere; the line list, providing the atomic (and molecular) data regarding the interactions between radiation and matter in the atmosphere; and the chemical composition, providing the distribution of the different elements in the gas and hence the ionisation equilibrium of atoms and the dissociation equilibrium of molecules. Moreover, chemical composition is key for calculating continuous absorption, scattering coefficients and partition functions.  
Usually,  the chemical composition of the Sun is adopted, in our work being that  of \cite{Grevesse2007}. This composition can then be scaled appropriately to different stellar metallicities, which are usually driven by the model atmosphere taken into account { and should be consistent with the composition of the model atmosphere}.  

However, the abundances of the so-called $\alpha-$elements (referring to O, Ne, Mg, Si, S, Ar, Ca and Ti) are different for different Galactic stars which are due to the different supernova Type II rates at different moments and parts of the Milky Way \citep[see][for a review]{2013ARA&A..51..457N}. The general trend is that at solar metallicities, $\alpha-$elements are solar-scaled ($[\alpha/\mathrm{Fe}] = 0$) and the $\alpha$-element abundance linearly increases towards lower metallicities, reaching a plateau of  $ [\alpha/\mathrm{Fe}] \sim 0.4$ at $[\mathrm{Fe}/\mathrm{H}] \sim -1$  \citep[see e.g.][]{1993A&A...275..101E}.

Non-solar abundance patterns can change the overall opacities and therefore the radiative transfer equations can lead to different solutions for the continuum flux as well as for abundances. Carbon-enhanced stars are a good example of this effect, as this element is very abundant  in stars, its interactions with photons occur over a large frequency range, and it contributes to continuous absorption in several ways.  Its impact in the model atmosphere has been discussed in \cite{Gustafsson2008}, in which continuous opacities of \ion{C}{I}, \ion{C}{II}, \ion{C}{$^{-}$}, as well as  CH and CO have been taken into account, showing to be significantly different for carbon-enhanced metal-poor stars. The enhancement  of $\alpha-$elements is more subtle than carbon. Although the effects in the temperature-pressure structures in model atmospheres are minor, for cooler stars can become significant, in particular due to TiO absorption. At the same time, electron contributions from Mg and Ca increase, raising \ion{H}{$^{-}$} opacities. The higher electron pressure causes a decrease in densities and thereby convective flux, which results in a stronger temperature gradient in the deep atmosphere. Therefore, special atmosphere models  of $\alpha-$enhanced stars are nowadays available to the community. This $\alpha-$enhancement in the MARCS models concerns continuous opacities of  \ion{O}{I}, \ion{O}{II}, \ion{O}{$^{-}$}, \ion{Mg}{I}, \ion{Mg}{II}, \ion{Si}{I}, \ion{Si}{II} , \ion{Ca}{I}, \ion{Ca}{II} as well as TiO.  In particular, the model grid employed in this work, as well as in the GES and the ASPCAP pipeline,  follows the general trend of the Milky Way, i.e. it does not take into account outlier stars such as  some accreted halo stars that are have $\alpha-$abundances lower than the plateau value of 0.4 mentioned above \citep{2010A&A...511L..10N, 2014MNRAS.445.2575H}, or higher $\alpha-$abundances at solar metallicities \citep{Adibekyan-12, 2015A&A...579A..52N}.

{ In the abundance analysis of $\alpha-$enhanced stars $\alpha-$enhanced model atmospheres should be employed together with consistent input abundances for the radiative transfer calculation. Unfortunately, in the literature the consistency in composition of the model atmosphere and the radiative transfer calculation is not always clearly documented. }Furthermore, it is not clear how the different radiative transfer codes compare when this scaling in the abundances of $\alpha-$elements is taken into account or when the abundances are assumed to be solar-scaled. 

In principle, this applies every element, but beyond C, N, $\alpha-$ and iron-peak elements, the number of other elements in a stellar atmosphere { are too small } to significantly affect the overall opacities and radiative transfer equation solutions.  In \fig{fig:alpha_syn} we  show the syntheses performed with \sme\ unad \ulb\ considering $\alpha-$enhanced and solar-scaled abundances for the 3 stars that are more metal-poor than the Sun, for Cr which is the bluest line analysed here, and for Mn, which is the reddest line. The syntheses were done without the inclusion of line absorption, and are intended to show the impact on the continuum flux and the continuum opacities due to $\alpha-$enhancement. 

\begin{figure}[!t]
\includegraphics[scale=0.6]{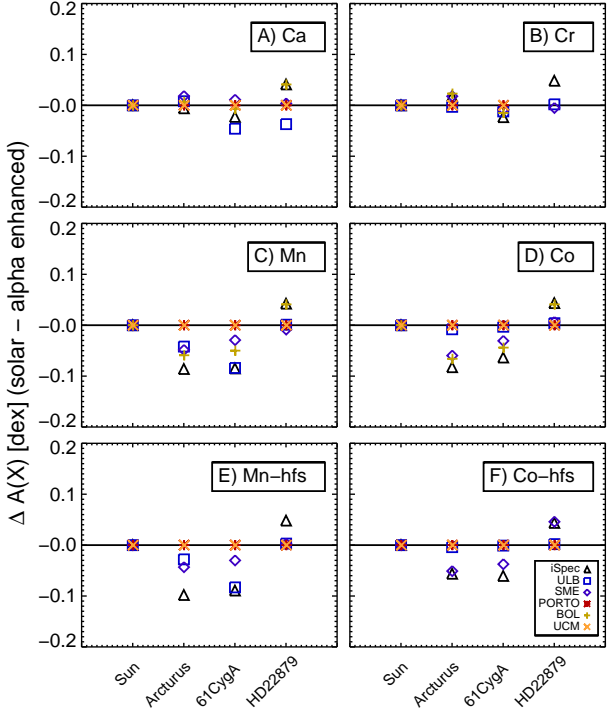}
\caption{Difference in element abundance derived by each method for different $\alpha$-element treatment. }
\label{fig:alpha}
\end{figure}

In each panel of \fig{fig:alpha_syn} we plot the absolute flux of the syntheses around a spectral window. The units of the flux are indicated in the y-axis. The mean difference between $\alpha-$enhanced and solar-scaled (in percentage) of the flux in the spectral window for each method is indicated in the legend.  We can see that although the same models are used for the synthesis computation, the treatment of opacities due to $\alpha-$elements is different for both radiative transfer codes, in which the continuum flux of \sme\  is systematically lower than the continuum flux of \ulb. Furthermore,  the differences between $\alpha-$enhanced and solar-scaled syntheses are systematically larger in the case of \sme\ with respect to \ulb, up to an order of magnitude for Arcturus.

As a next step we used our methods to estimate typical uncertainties in derived abundances due to this effect. Thus, we performed a test in which the abundances were derived twice as follow:

\begin{enumerate}
  \item each method determined abundances scaling the abundances according to the standard $\alpha-$enhancement of the Milky Way  ($\alpha-$enhanced),
   \item each method determined abundances considering solar-scaled abundances (solar).
\end{enumerate}

In both cases, the same model-atmosphere was used, i.e, following the standard $\alpha-$enhancement as a function of metallicity of the Milky Way. 
In \fig{fig:alpha} we show the results obtained in this test. Each panel corresponds to an element, in which Co and Mn were determined with and without HFS. Each method is represented by a different symbol, which is the same throughout all figures. The differences between the results obtained considering solar abundances and $\alpha-$enhanced are shown for each star. { For the Sun the abundance difference is zero, since $[\alpha/\mathrm{Fe}] = 0$ in both cases.}

For more metal-poor stars, { differences in abundances of up to 0.1~dex were found when comparing the two cases of chemical composition}  except for the methods using MOOG (\porto\ and \ucm).  MOOG itself  does not seem to calculate the electron number density, it rather  uses the values read from the model atmosphere throughout its calculations. Since  the model atmospheres were not changed in the test, the electron number density did not change either, and thus the continuous \ion{H}{$^{-}$} opacity is the same.

We verified that all other radiative transfer codes used in this work ({\tt sme\_synth}, Turbospectrum, SPECTRUM, WIDTH9) do include their own computation of the electron number density (or electron pressure) while solving the equilibrium equations, taking into account the electron contributions from the ionisation of different species\footnote{See e.g. SPECTRUM manual p. 33 (and file density9.c); WIDTH9/ATLAS9 subroutine NELECT; Turbospectrum file tryck.f}. This is not necessarily a problem for the abundance calculations with MOOG, since the model atmospheres include $\alpha-$enhancement and the corresponding electron pressure. It means that MOOG does not need to adjust the abundances, but all others should do so, to be consistent with the model atmospheres. 

Interestingly, the most metal-poor star HD~22879 which should be the most $\alpha-$enhanced star of the sample does not show the largest difference in results. The cool stars Arcturus and \cygA\ show a systematic difference of abundances being  higher for $\alpha-$enhanced composition with respect to solar. { This is probably due to the lower continuum flux obtained for the alpha-enhanced composition (\fig{fig:alpha_syn}).
However, the differences of the metal-poor warm star have a different sign than the cool stars. This suggests that the changes in pressure structure affect the abundances in the opposite direction than the changes in continuum opacities.}

\subsubsection*{Conclusion for $\alpha-$enhancement and continuum opacities}

When $\alpha-$abundances are different of solar for stars, it is important that the abundances are consistent with the model-atmosphere. While MOOG does it automatically, this is not the case in most of the classical methods to determine abundances.   If the abundances are not accordingly scaled, differences of up to 0.1~dex can arise in retrieved abundances of cool stars.  This uncertainty is larger than aimed for   current pipelines to determine abundances in large spectroscopic surveys. { We recommend to always specify the chemical composition adopted in the radiative transfer calculations when publishing abundances. The size and sign of the abundance difference seems to vary stellar parameters in a complex way, depending on the relative importance of opacity and pressure effects. Further studies for a wider variety of stars and transitions are needed to assess this issue in more detail.}

\begin{figure*}[!t]
\includegraphics[scale=0.45]{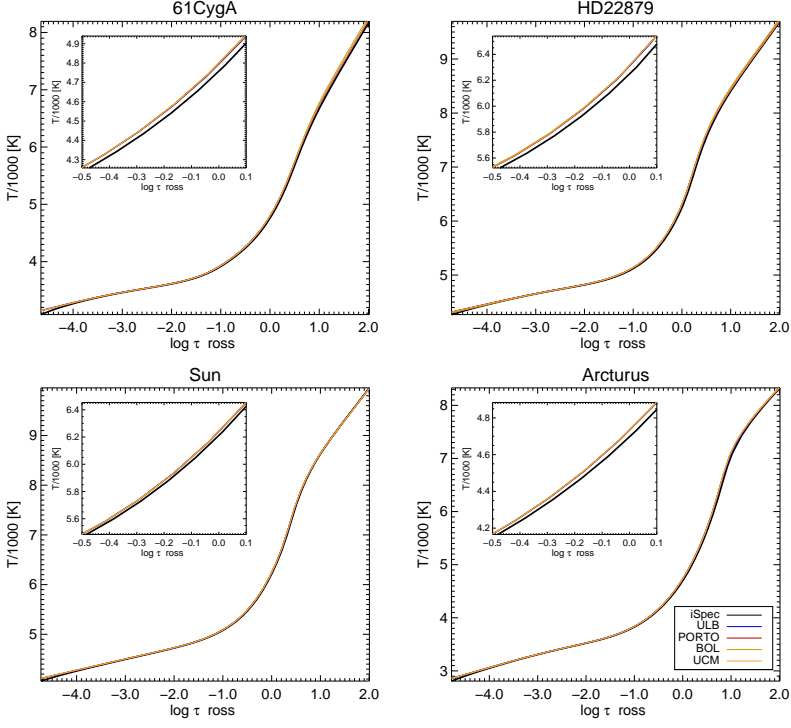}
\vspace{-0.3cm}
\caption{{ Comparison of interpolated model temperatures as a function of Rosseland optical depth. The profiles for all methods except for \ispec\ lie on top of each other.}}
\label{fig:model_inter_profiles}
\end{figure*}

\begin{figure*}[!t]
\includegraphics[scale=0.45]{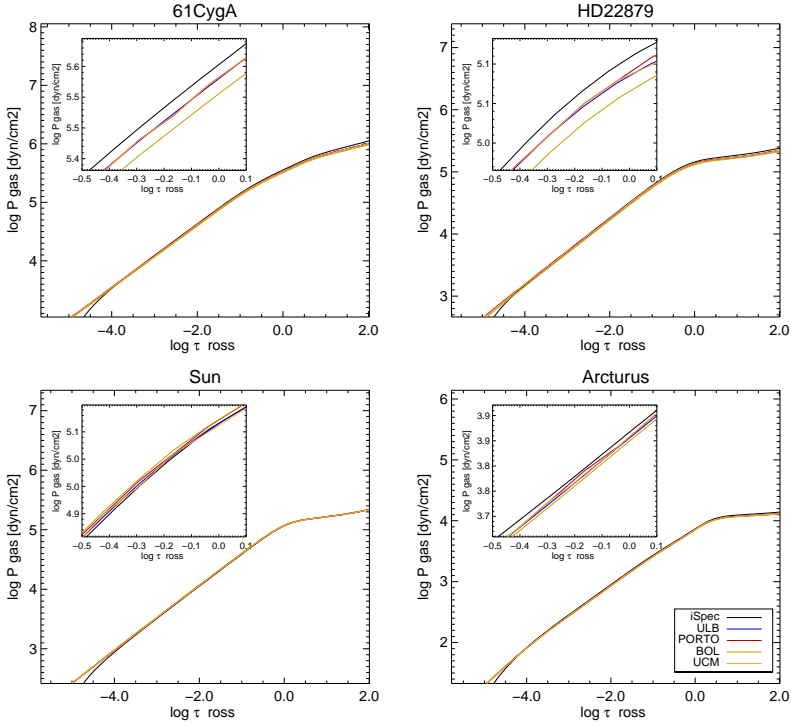}
\vspace{-0.3cm}
\caption{{ Comparison of interpolated model gas pressures as a function of Rosseland optical depth.}}
\label{fig:model_inter_profiles_P}
\end{figure*}

\subsection{Atmospheric model interpolation}

When deriving stellar parameters and abundances methods require an appropriate model atmosphere, which usually is taken from a grid of models, like the MARCS one in this case. Other widely-used 1D-LTE model atmosphere grids are created with ATLAS \citep{1979ApJS...40....1K} and PHOENIX \citep{2013A&A...553A...6H}. More sophisticated model grids considering 3D have also become available recently like the Stagger grid \citep{2013A&A...557A..26M}. These grids cover the parameter space at fixed steps in \teff, \logg\ and \feh, with the 3D grids having larger steps than the 1D grids as the models are computationally more expensive to generate. In either case, it is very common that when deriving abundances the model atmosphere computed for the exact parameters of the star is not available, so the radiative transfer code operates considering an {\it interpolated } model atmosphere of the exact stellar parameters. How this interpolation is done is rarely documented in spectroscopic methods. Here, \ulb\ and \bol\ follow the interpolation procedure of \cite{MasseronThesis}, which is the same as the interpolation procedure adopted by the MARCS grid described in \cite{Gustafsson2008}.  \porto\ and \ucm\  employ interpolation code available on the MARCS models web page, and therefore these models should be essentially the same as those of \bol\ and \ulb. \sme\ is interpolating the model atmospheres in a very similar way to \cite{MasseronThesis} with a trilinear interpolation
{ between eight corner models surrounding the target parameters. The procedure has three steps, first obtaining four models at the requested \feh, followed by two models with the desired \logg, from which a single output model is produced by interpolation in \teff.}
On the other hand, { \ispec\ first does a bilinear interpolation in \teff\ and \logg\ between neighbouring models, followed by a linear interpolation in  \feh.}

{ In order to compare interpolated models we first transformed the model data produced by each method onto a common depth scale, where possible. }
 \ulb\ provided both the optical depth at $500~\mathrm{nm}$ ($\tau_{500}$) and the Rosseland optical depth ($\tau_{\rm ross}$), while \sme\ included only $\tau_{500}$. The remaining methods provided the Rosseland opacity $\kappa_{\rm ross}$ and the mass variable $\rho dx$, which were converted to $\tau_{\rm ross}$ by integration. Further conversions involved those between number densities and pressures, and between quantities given in linear or logarithmic form.  
 
 In \fig{fig:model_inter_profiles} we show the  interpolated model profiles of all methods with different colours. For this illustration we show the temperature ($T$) as a function of $\tau_{\mathrm{ross}}$.  Each panel indicates a star, and the sub-panel shows a zoom of the model around $\log \tau_{\mathrm{ross}} \sim 0$. The models produced by \sme\ are not included in this figure, but were compared with those of \ulb on the $\tau_{500}$ scale and found to be identical. Here we define equal when models differ by less than $0.1\%$. Therefore,  we assume that the results from the comparison shown in \fig{fig:model_inter_profiles} can also be applied for the models interpolated by \sme.

From \fig{fig:model_inter_profiles} we can see that the models of \ulb, \bol, \ucm\ and \porto\ are equal while at $\tau_{\mathrm{ross}} \sim 1$  \ispec\ is slightly different with respect to the rest of the models.   
For the Sun we find differences of $-30~\mathrm{K}$ for $T\sim6230~\mathrm{K}$ corresponding to $0.5\%$,  for Arcturus a difference of  $-40 \mathrm{K}$ for $T\sim4730 \mathrm{K}$ corresponding to $0.8\%$, for HD22879 we see a difference of $-60~\mathrm{K}$ for $T\sim6310~\mathrm{K}$ corresponding to $1\%$ and for \cygA\ the difference is of $-40~\mathrm{K}$ for $T\sim4800~\mathrm{K}$ corresponding to $0.8\%$.

The gas pressure  is plotted as a function of  $\log \tau_{\mathrm{ross}} $ for all models except \sme\ in \fig{fig:model_inter_profiles_P}. As in previous comparison, we found that \ulb\ and \sme\ are the same and therefore we can assume that the comparison of \sme\ with respect to the rest of the methods is the same as the comparison of \ulb\ with respect to the rest of the methods. In contrast to the temperature profile, we see that there are more significant differences in the line-forming region between the different interpolated models.  For the Sun we see a difference of 2\% to 5\%, for Arcturus of 5\%, for HD22879 and \cygA\ of 10\%.

\begin{figure}[!t]
\includegraphics[scale=0.6]{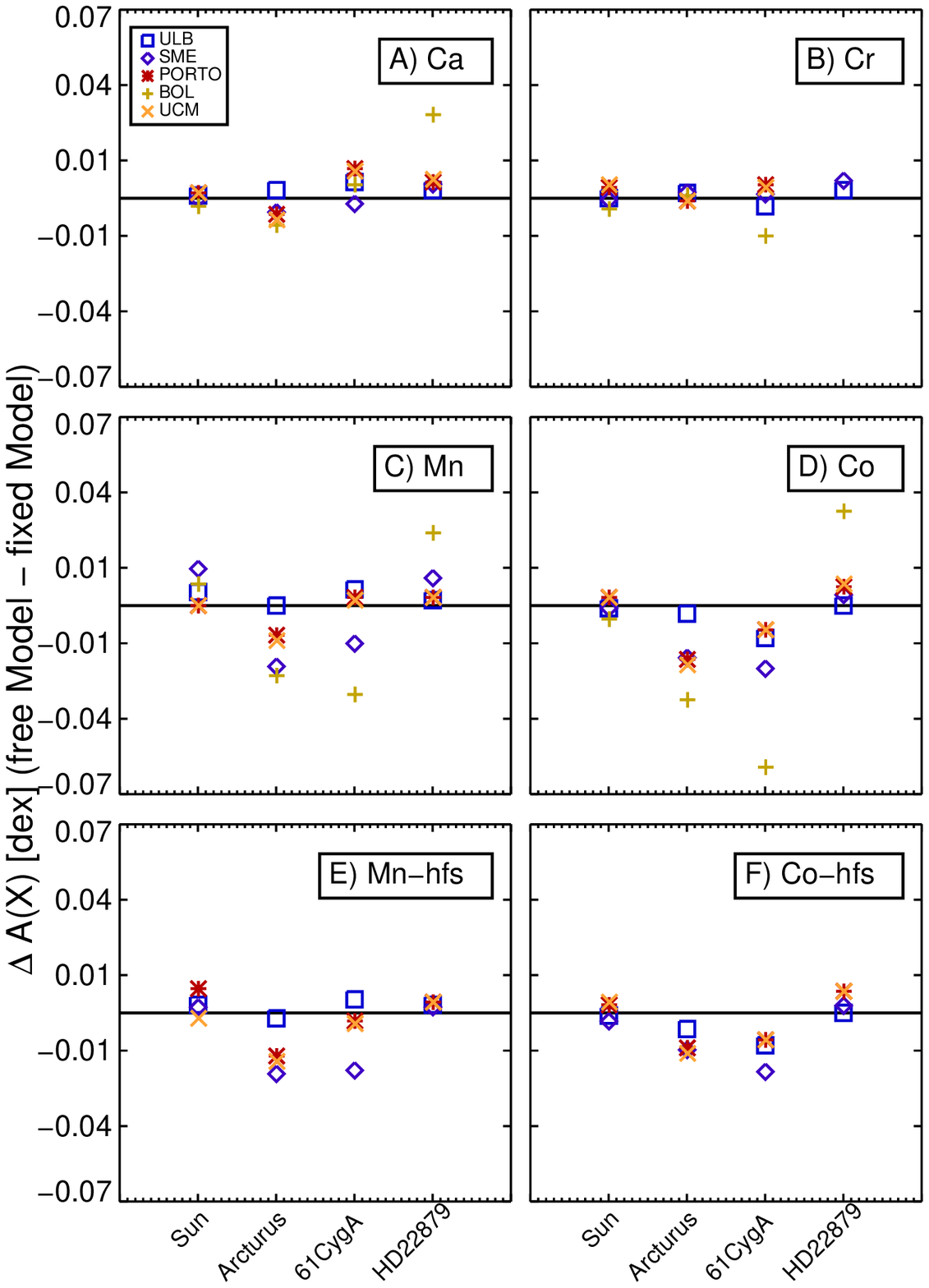}
\vspace{-0.8cm}
\caption{Difference in element abundances derived by each method by performing own interpolation of model atmospheres or considering a pre-defined model. }
\label{fig:model_inter}
\end{figure}

Here we estimate typical uncertainties in abundances due to different interpolations of atmosphere models. To do so, the abundances  were determined two times as follows: 
\begin{enumerate}
  \item the methods determined abundances performing own interpolation of model atmospheres, i.e, using the model profiles shown in \fig{fig:model_inter_profiles} and \fig{fig:model_inter_profiles_P} ($free\_model$), 
   \item the methods determined abundances using a fixed model ($fixed\_model$). 
\end{enumerate}
The $fixed\_model$ for each star was created using \ispec. The reason is that the latest version of \ispec\ \citep[Blanco-Cuaresma et al. {\it in prep.}, see also ][for further details]{2016arXiv160908092B} has now included all  1D LTE radiative transfer codes used in the present work, namely {\tt sme\_synth}, MOOG, Turbospectrum, SPECTRUM and SYNTHE9. Therefore, with \ispec\ it is possible to store the same interpolated model into different formats that can be then applied by the different methods considered in this work.  Furthermore, the interpolation procedure of \ispec\ differs more from the others, making the comparison of retrieved abundances more significant in this case.

\begin{figure}[!t]
\hspace{-0.6cm}
\includegraphics[scale=0.55]{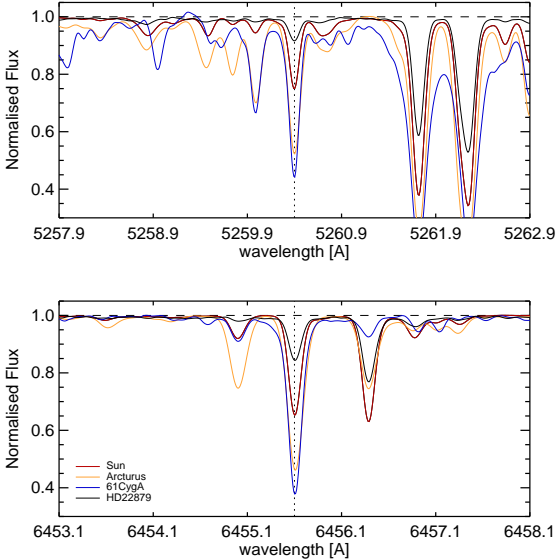}
\caption{Observed profiles of two Ca lines in the four GBS. The upper panel shows the profiles for the Ca line analysed throughout this work, while the lower panel shows the profiles for a cleaner Ca line }
\label{fig:7profiles}
\end{figure}

In \fig{fig:model_inter} we illustrate the results of this test. In each panel we show the difference obtained between using the pre-defined atmosphere model of \ispec\ and performing own interpolation. Because of the construction of this test, we do not plot the \ispec\ results as they are the same in both cases. For the other methods, we see that typical differences are of less than 0.04~dex for all stars and lines. { Although the interpolated temperature structures were found to be very similar for all methods, the abundance differences vary considerably. } For example, while \ulb\ is very little affected by the input model in most of the cases, \sme\ and \bol\ seem to be most affected, in particular for the metal-poor and the cool stars. 

 The methods however respond to the same direction in almost all cases, showing that the dependency of the model when solving the radiative transfer equations is consistent among methods. The magnitude of the dependency is also consistent with the general difference obtained between methods, i.e. the largest difference in the gas pressure versus $\tau_{\mathrm{ross}}$ was obtained for 61~Cyg~A of 10\%, which shows to have the largest difference in abundances obtained by all methods.  { We comment here that this test was done for stars with parameters that are well inside the grid of atmospheres. The impact on abundances obtained for stars with parameters on the edge of the grid was not explored since the stars used in this work have parameters that lie well within the grid of models. The extrapolation of atmosphere-models might cause larger differences in abundances than found here.  }

  { A potential test which was not performed here but is of similar nature would be to study the difference of abundances derived when standard model atmospheres or smodels with self-consistent chemical composition are used. For example,  the influence of Mg, Al, Na as electron donors or C in molecular equilibrium can be significant on the structure of the atmosphere of a red giant. This can affect the abundances derived in e.g. second generation globular cluster stars where the $\alpha-$elements O, Mg  and C are depleted whereas Al and Na are enhanced. To quantify this effect would require model atmospheres calculated for the exact compositions of second generation globular cluster stars. }

\subsubsection*{Conclusion on model interpolation}

Even if models are interpolated following very similar approaches, subtle differences of about 1\%  in the temperature and of 5\% in gas profiles are found. This is translated to differences of the order of 0.02~dex in the derived abundances, although this number depends on the method considered. Such differences are of the order of magnitude of uncertainties accepted by big surveys, but are larger than those aimed for very high precision abundance studies such as those of \cite{2015A&A...579A..52N} or \cite{2016arXiv160604842S} of 0.01~dex.

\begin{figure}
\includegraphics[scale=0.6]{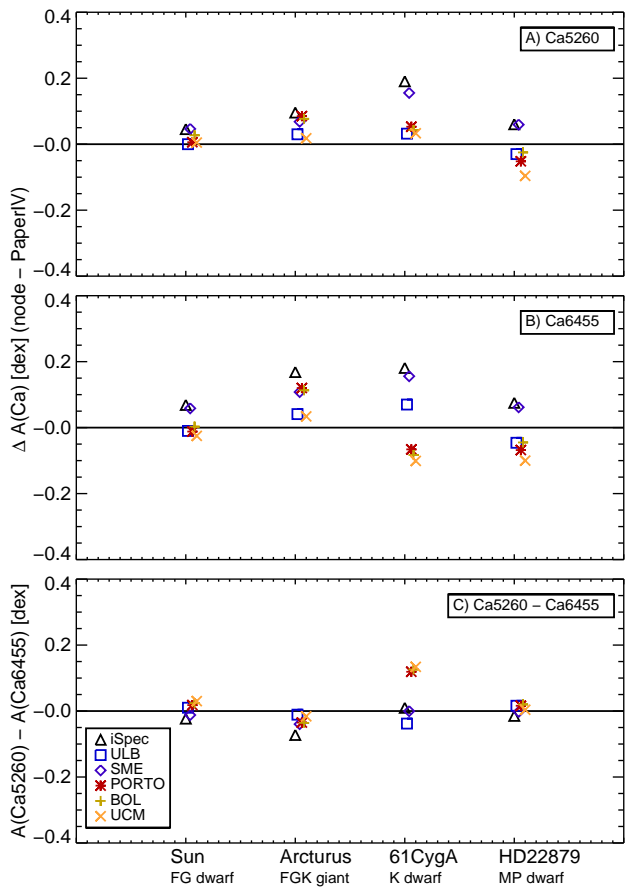}
\vspace{-1.5cm}
\caption{Panel A and B: Differences  in Ca abundances with respect to a reference value (Paper~IV) obtained from two lines with different blend levels.  { Panel A shows the results of the more blended line. Panel C shows the differences between the abundances from both lines for each node. }  
}
\label{fig:blends}
\end{figure}

\subsection{Blends}\label{blends}

Although abundance determination from spectral lines aims at analysing lines that are free of blends and isolated enough to identify robust continuum points, it is difficult to find perfect clean lines for all type of stars (see also extensive discussion in Paper~IV).  Indeed, even in our careful choice of optical lines for this work, it was not possible to find perfectly clean and unblended lines for all 4 different GBS. The example of the Ca line is seen in the upper panel of \fig{fig:7profiles}, which shows the region around the \ion{Ca}{I}~$\lambda 5260$~\AA\ line. One can see that the right wing of the line is blended, even in the metal-poor star HD22879. This blend corresponds to a blend of \ion{Si}{I} and \ion{Mn}{I}. The lower panel of \fig{fig:7profiles} shows another \ion{Ca}{I} line ($\lambda 6455$~\AA), which is placed at a less crowded region and does not show strong blends on its wings, except for the left wing of \cygA. 

Although synthesis methods can cope better with blends by disregarding the blended regions and synthesising only part of the lines, this is not always done as these blends need to be previously identified, with the  amount of blend varying from star to star. The EW methods suffer more evidently from this effect as the total area covered by the line is larger than the actual area filled by the corresponding element, yielding larger abundances \citep{Stetson2008}.

 In this test we attempt with our methods to estimate typical uncertainties in abundance of a given element due to different blend levels in different lines. To do so, the abundances of Ca were determined two times as follows: 
\begin{enumerate}
  \item each method analysed the \ion{Ca}{I} line $\lambda 5260$~\AA. 
   \item each method analysed the \ion{Ca}{I} line $\lambda 6455$~\AA. 
\end{enumerate}

The atomic data for the  \ion{Ca}{I} line $\lambda 6455$~\AA\  are from the same source as for the \ion{Ca}{I}  $\lambda 5260$~\AA\ line \citep{SR}.

\begin{figure}[!t]
\includegraphics[scale=0.6]{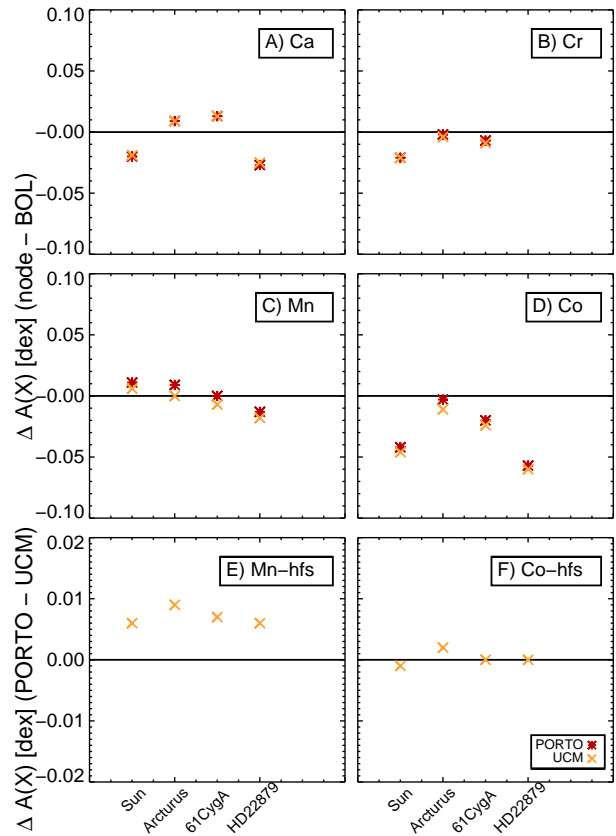}
\vspace{-1.2cm}
\caption{Abundances obtained for same EW and different radiative transfer codes. Only methods using EWs (\bol, \ucm\ and \porto) are taken into account. These methods use the radiative transfer codes SYNTHE (\bol) and MOOG (\ucm\ and \porto).  }
\label{fig:8.same_ew}
\end{figure}

In \fig{fig:blends} we illustrate the results of this test, where we plot the difference between the results of each method and the value obtained for the line in Paper~IV as reference.   Panel A shows the results obtained for the $\lambda 5260$~\AA\ line and Panel B the results obtained for the $\lambda 6455$~\AA\ line. In Panel C we show the difference in the abundances obtained from each line.  

We do not find a significant difference in the node-to-node scatter for a the more blended $\lambda 5260$~\AA\ line with respect to the less blended  $\lambda 6455$~\AA\ line. The larger scatter among methods is obtained for cooler stars for both lines. The differences obtained between both lines are in general very small for all methods and stars except for the three EW methods for \cygA\ with a difference of the order of 0.15~dex. { This suggests that abundances derived from EWs are only affected in the most extreme case of blending (see \fig{fig:7profiles}), while the synthesis method is robust with respect to blends in all cases. On the other hand, the method-to-method scatter in absolute abundances for 61~Cyg~A is actually smaller for the more blended line (\fig{fig:blends}, panels A and B), indicating that the effect of the blends is working in the direction opposite to other differences between methods.}


\subsubsection*{Conclusion for blends}
{ From this test we conclude that synthesis and EW methods result in similar abundances for lines with moderate blend levels. The \ion{Ca}{I} $\lambda 5260$~\AA\  line may serve as an example for the limiting case.} The differences obtained from method to method is therefore dominated by other uncertainties, such as those discussed throughout this paper. {  It is worth to comment that this conclusion is obtained from this test only, applied to these two lines in these 4 stars. See also \cite{2015A&A...583A..94A} for further discussions on this matter.}

\subsection{Radiative transfer codes}

In this work we employ six different methods that can be divided into two different fundamental methods: equivalent widths and syntheses. The differences among EWs methods can be due to a difference in the measured EW, either by continuum placement, or by the line fitting procedure, among others. The differences among syntheses can also be due to different values adopted for the instrumental resolution, the macroturbulence parameter \vmac\ parameter or the specific spectral region to synthesise. If the aforementioned quantities are fixed, then one can explore the differences obtained in abundances due to the radiative transfer codes, which is what we explore in this section. For that, we separate the analysis into these two fundamental types, namely EWs and syntheses.

\begin{figure*}[!t]
\includegraphics[scale=0.8]{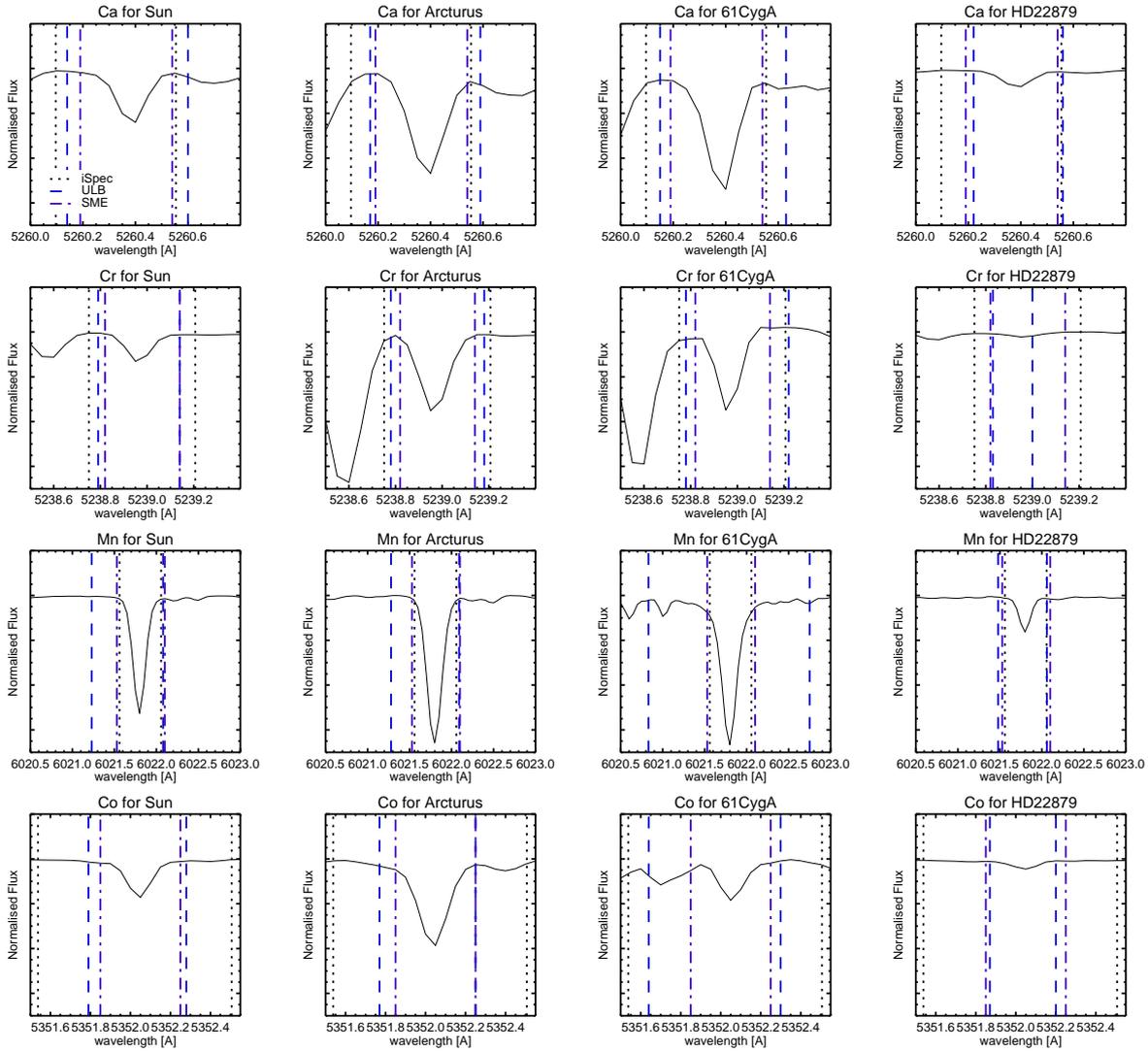}

\caption{Line regions employed by the three synthesis methods considered in this work are enclosed by the vertical lines of the corresponding colour. Line profile of each star and element is also shown for reference. }
\label{masks}
\end{figure*}

\subsubsection{Equivalent widths}

In this section we explore the differences of the radiative transfer codes when the same EW is assumed. More specifically, we compare the results obtained by \bol\, \ucm\ and \porto. The radiative transfer codes to compare are SYNTHE as used by \bol\ which calls this code from GALA \citep{2013ApJ...766...78M}; and MOOG as used by \ucm\ with the code {\tt StePar} \citep{2012A&A...547A..13T} or by \porto\ as described in Paper~III and Paper~IV. 

For this comparison, we fixed the EWs of \bol, which have been used for all the tests of this paper except those presented in \sect{2.Cont_norm} for the continuum normalisation test.  In \fig{fig:8.same_ew} we show the differences obtained in the abundances of all elements between \bol\ and the other 2 EWs methods. In other words, the figure shows the difference between SYNTHE and two versions of using MOOG. For the bottom panels E and F { we show the differences between \porto\ and \ucm, to see the difference under HFS.} 

In general we can see that the abundances derived by \ucm\ and \porto\ are more similar to each other than when compared the abundances with \bol. This is not surprising as \ucm\ and \porto\ use the same radiative transfer code, MOOG. For Ca and Cr (Panels A and B of \fig{fig:8.same_ew}) { the differences between \ucm\ and \porto\ are negligible for most of the stars, while the differences of them with respect to \bol\ are less than 0.05~dex.  For Mn and Co without HFS (Panel C and D of  \fig{fig:8.same_ew})  the differences between \bol\ and \ucm/\porto\ can reach up to 0.06~dex, as in the case of Co for the metal-poor star. Even when the same radiative transfer code (MOOG) is used, relative differences of 0.01~dex appear in the case of Arcturus for Mn and Co with or without HFS (panels C to F). } This hints towards other differences in how the codes are performing the fitting by the different methods rather than the radiative transfer code itself.  

We finally note that the weakest lines (those of the Sun and HD 22879) have the largest difference between  SYNTHE and MOOG results. { This can be explained by the high sensitivity of the line strength of weak lines to abundance, i.e. the steep slope of the curve of growth at the weak-line end.}

\subsubsection{Syntheses}

\begin{figure*}[!t]
\includegraphics[scale=0.5]{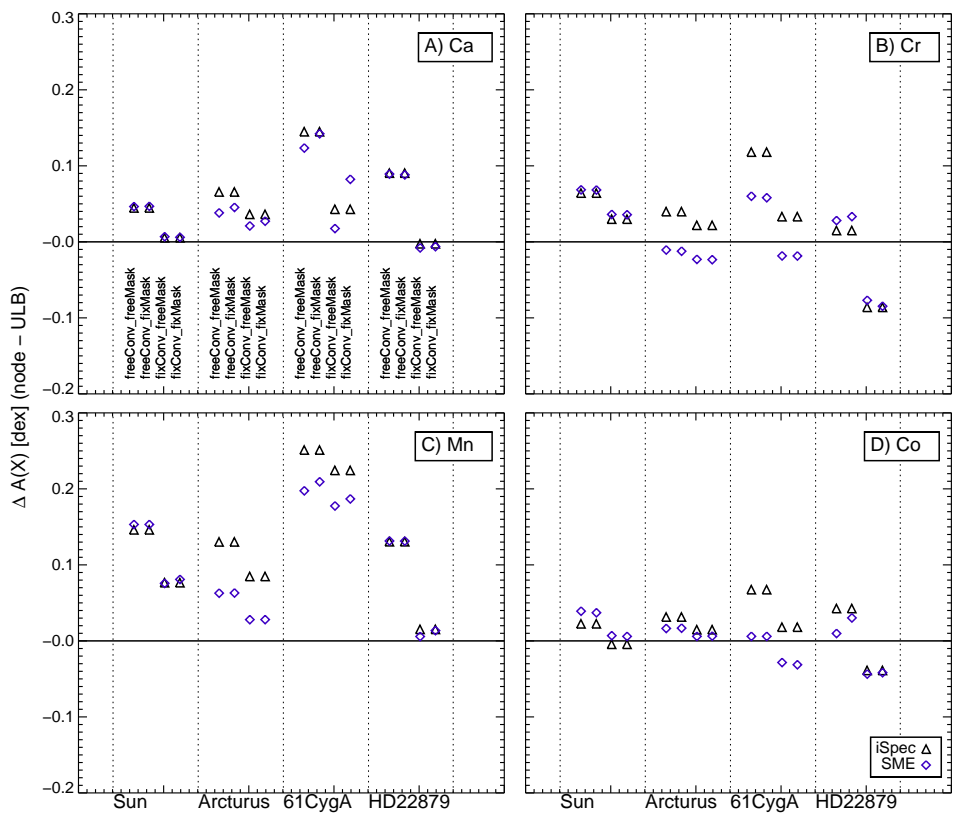}
\vspace{-1cm}
\caption{Abundances obtained for same masks/convolutions but different synthesis codes. Only methods using synthesis (\ulb, \sme\ and \ispec) are taken into account. These methods use the radiative transfer codes Turbospectrum (\ulb), sme\_synth (\sme) and SPECTRUM (\ispec).}
\label{fig:same_syn}
\end{figure*}

 In this test we attempted, with our methods, to estimate typical uncertainties due to different  radiative transfer codes when synthesis methods are considered.  We compare the results obtained with the methods \ulb, \sme, and \ispec, which use the radiative transfer codes Turbospectrum \citep{2012ascl.soft05004P}, sme\_synth \citep{1996A&AS..118..595V} and SPECTRUM \citep{1994AJ....107..742G}, respectively. An equivalent to fix the EW of the previous section in this case would mean to fix the spectral region (mask) where the fitting is performed for the abundance determination.  In addition, each method performs a convolution to  model different broadening effects in a line profile. In this section we explore these two parameters.
 
We first compare the masks (regions) considered by the three methods in each case. This is shown in \fig{masks},  where in each panel we plot the line profile of a different star and element. The masks are represented as vertical lines with different colours and lines for each method. 

In general the regions agree very well within the 3 methods, with few exceptions such as the masks for Co of \ispec\ or the Mn mask for \cygA\ of \ulb. While \ispec\ employs the same mask for a given element, \ulb\ and \sme\ adjust the mask according to the properties to the line profile of each star individually. This procedure is also possible for \ispec\ in other applications \citep{2015A&A...577A..47B}, but was not implemented in this work.  The criteria to adjust masks is clearly different between \sme\ and \ulb, { obtaining sometimes significantly different regions for performing the fitting}, such  as Mn of \cygA\ or the weak Cr line  of HD22879.  We remark that the methods have independently developed a procedure to choose masks, but the final outcome is in most of the cases very similar, i.e. less than 1\AA\ difference.

 Regarding broadening parameters, these account for a better reproduction of the instrumental profile, which is usually  treated as a gaussian smoothing of the synthetic spectrum to match the resolution of the observed spectrum.  Line broadening can also be  caused by stellar rotation, macroturbulence and other effects.  { Unlike microturbulence, these broadening mechanisms do not change the strength of a spectral line.} In \ispec\ and \sme,  rotational broadening and macroturbulent broadening  are applied  after  the synthetic spectrum has been convolved with the appropriate line-spread functions given by the instrumental resolution.
 
While \sme\ and \ispec\ need the three parameters (\vmac, projected equatorial velocity \vsini\ and the resolution $R$) as independent input for the computation of synthetic spectra, \ulb\ employs only one   convolution parameter, while \sme\ and \ispec\ considered $R=70,000$,  and \vsini\ and \vmac\ as listed in \tab{t:bs_ap}, \ulb\ considered an overall convolution parameter of 5.8, 8.0, 4.6 and 5.9 \kms\ for the Sun, Arcturus, \cygA\ and HD22879, respectively.  

To study the effect of both, broadening and mask choice, in the final results we determined the abundances four times as follows: 
\begin{enumerate}
  \item each method determined abundances with the same convolution and fixed mask ($ fixConv\_fixMask$),
   \item each method determined abundances with the same convolution and free mask ($ fixConv\_freeMask$),
   \item each method determined abundances with a different convolution and fixed mask ($ freeConv\_fixMask$),
   \item each method determined abundances with different convolution and free mask ($ freeConv\_freeMask$).
\end{enumerate}
Here HFS was considered for the analysis of the Mn and Co lines.  The masks and the convolution parameters were always fixed by \ulb, and therefore our analysis consisted in comparing the results of \ulb\ with \sme\ and \ispec. Thus, for  the fixed tests,  \vmac\ and \vsini\ were considered to be zero by \ispec\ and \sme, and the resolution was considered to correspond to the convolution parameter of \ulb, { which corresponds to $ R = c/v_{\mathrm{ULB}} = 51688, 37474, 65172$ and $50812$ } for the Sun, Arcturus, \cygA\ and HD22879, respectively.  When convolution or the mask were  free, it meant that \sme\ and \ispec\ used their own parameters.

In \fig{fig:same_syn} we show the results of this test. In each panel we show one element, and each star is separated by vertical dashed lines. The four tests are indicated for each star in each panel, following the same order as in Panel A. We compared the results of \sme\ and \ispec\ with respect to \ulb\ in each case. 

{ When comparing the $freeMask$ with the $fixMask$ cases}, we see that \ispec\ obtains essentially the same results. The same happens for most of the masks adopted by \sme, with notable exceptions being the the Ca and Mn masks for \cygA, which  differ more significantly (see \fig{masks}).  This test is telling us that the methods are very robust in the abundance measurement when different masks are employed. A notable case is Co for \ispec, presenting negligible differences when two very different masks are considered. This might be due to the fact that results will not be affected if regions not sensitive to the abundance are considered in the fit.

A more notable difference is obtained when the broadening parameters (convolution) differs. Both \sme\ and \ispec\ show very similar differences to \ulb, but we remind that both methods use the same parameters.  There is a general tendency of decreasing the final abundance when the \ulb\ convolution parameter is taken into account. This suggests that the overall convolution of \ulb\ has a larger value than  the total convolution of the other two methods. A broader line will have a weaker core, imitating a lower abundance. This is because the core of a weak line -- lines with EWs of about $120~\mathrm{m\AA}$ { (that is, a reduced EW of  about $\log{(\mathrm{EW}/\lambda)} = -4.8$ in the optical, see Paper~III)} or less --  is usually the region with the largest sensitivity to the abundance and where most of the datapoints to perform the fitting lie.  We see that in both cases the increase in broadening is translated to a decrease in abundance. This input difference drives differences of the order of 0.05~dex in abundances for most of the cases, but could reach more than 0.1~dex in some cases like Mn for HD~22879.

Finally, we see that even adopting the same masks and convolution parameters (rightmost points for each star in \fig{fig:same_syn}), the differences between \ulb\ and \sme\ or \ispec/ are not fully alleviated. While in some cases this indeed happened, like the abundance of Ca for the Sun and HD22879, this did not happen for the same stars  for Cr, for which the differences increased.  We might attribute this uncertainty to an intrinsic difference between Turbospectrum, SPECTRUM and sme\_synth, or to some other effect such as the fitting procedure. For \sme\ and \ispec, syntheses of a given line are computed on-the-fly for different abundances values and a $\chi^2$ is calculated between the synthetic and observed profile. If the $\chi^2$ is higher than the desired convergency parameter, new syntheses are computed considering abundances values towards the direction in which $\chi^2$ was better. This process is repeated until convergency is achieved. 
The fitting procedure of \ulb\ is slightly different. Only five syntheses are computed with different abundance values and a $\chi^2$ is calculated for each of them. A likelihood function is then obtained, and the solution is found via interpolation of that function. 

To check if this different approach produces discrepancies of up to 0.1~dex  one would have to modify the fitting procedures of the codes, which goes beyond the scope of this work. 

\subsubsection*{Conclusion for Radiative transfer codes}
When { the } same EWs are used, differences of up to 0.05~dex can be obtained when different radiative transfer codes are used. For lines affected by HFS, differences are more notable, even if the same radiative transfer code is employed. We conclude that the treatment of HFS is an important driver of differences between same radiative transfer codes, which can reach 0.18~dex for Mn.  

The choice of mask had in most of the cases very little impact in the determination of abundances, even if the masks differed  significantly. On the other hand, the convolution parameters can produce differences of up to 0.1~dex.

\begin{table*}[!t]
\caption{Summary of tests and uncertainties. For each test we list the typical { (median)} uncertainty obtained from all results in a given test (expected $\sigma$), as well as the maximum difference (Max $\sigma$) and the minimum difference (Min $\sigma$).} \label{t:summary}
\begin{tabular}{lccccc}
\hline\hline \\
Test & Expected $\sigma$  & Max $\sigma$ & Min  $\sigma$  & comment\\
        &  dex   &  dex &  dex    \\
\hline \\
1. Line list & { 0.05}  & 0.6 & 0.0 & EW are not affected by position of core of line \\
&&&& but number of HFS components { might be} important \\
\hline \\
2. Continuum & 0.3 & 0.6 & 0.05 & absolute abundance is very dependent \\
 & & & & but methods should not re-normalise\\
 \hline\\
3. HFS & { 0.08} & { 0.4} & { 0.0 } & { differences in abundances between HFS:N and} \\
&&&& { HFS:Y are similar for all methods} \\
\hline\\
4. \vmic &  0.2 & 1.2& 0.01 & { maximum difference corresponds to} 1 km/s range in \vmic, \\
&&&& dependency decreases when HFS is considered \\
\hline\\
5. $\alpha$-enhancement & 0.02& 0.1&0.001& cool stars are more affected than\\
&&&& metal-poor warm stars \\ 
\hline\\
6. Atmosphere model interpolation & 0.01 & 0.04 & 0.002 & obtained from models with 1\% difference in \\
&&&& temperature and 5-10\% difference in gas pressure. \\
\hline \\
7. Blends &  { 0.02} & { 0.1} & {0.0} & { maximum difference is found for }\\
&&&& { EW methods for cool dwarf star} \\
\hline \\
8.1. Same EWs & 0.02 & { 0.06} & 0.001 & weak lines are more affected\\
8.2. Same syntheses & 0.07 &{ 0.12} &0.001& convolution is more important than the choice of mask \\
\hline
\end{tabular}
\end{table*}

\section{Summary and concluding remarks}\label{concl}

 The goal of this work was to explore the impact on the derived stellar abundances produced by some of the technical assumptions of different methods commonly used in the field. For this, participants with expertise in the methods employed in our previous works on spectral analysis of GBS (Paper~III and IV) were considered. In this work we discussed the differences in abundances obtained by six methods, three methods based on equivalent widths (\bol, \ucm\ and \porto) and three methods based on syntheses (\ulb, \sme\ and \ispec).

We analysed four GBS, each of them being a reference star of a group of GBS as defined in \cite{Jofre2015}: The Sun representing \fgdwarf, Arcturus for \fgkgiant, \cygA\ as one of the \kdwarf, and HD22879 representing \metalpoor\ stars.  For each star, we determined the abundances of four elements, using one line each of Ca, Cr, Mn and Co. These lines were selected from  \cite{Jofre2015} because they were analysed by several methods for these stars. 

By fixing the stellar parameters and using the same { observed} spectrum, atmosphere model and line list, we performed 8 different tests with the goal to explore potential  sources of systematic uncertainties in the derived abundances. These are non-exhaustive and further tests should be explored in future works on this matter.   A summary of these tests with the main conclusions is given below and in \tab{t:summary}.  In the table, we list each  test and give a quote of the typical uncertainty considering all methods and stars, as well as the minimum and maximum uncertainty. These uncertainties are the median, minimum and maximum difference of all results analysed in the given test. 

\begin{enumerate}
\item {\it Linelist:} We investigated the impact on deriving abundances of Mn using two line lists, one of them having the core of the line slightly shifted with respect to the observations. Both line lists also had different binning for HFS. We found that when HFS was not taken into account, the EW methods showed no difference, while the synthesis method showed a difference of up to 0.6~dex. When HFS was taken into account, typical differences of { 0.05}~dex were found { in the case of syntheses, while of less than 0.01~dex for the case of EW methods}. \\

\item {\it Continuum normalisation}: We investigated the impact on deriving abundances of all elements considering different approaches for normalising the spectra. We took 2 spectra normalised by two methods and determined abundances by fixing the continuum or allowing the methods to re-normalise. We found that the method-to-method scatter decreased by up to 0.5~dex when the continuum was kept fixed. We also found offsets of 0.5~dex (Co for \cygA) in the final absolute abundances. We conclude that the continuum placement is key for an absolute result of the retrieved abundances, and that when combining different methods, the agreement improves if the continuum is kept fixed. \\

\item {\it Hyperfine structure}: We compared the results obtained for the abundances of Mn and Co considering and neglecting HFS. We found that differences of up to { 0.4~dex} can be obtained if HFS of Mn is not considered  while of 0.1~dex if HFS of Co is not considered. While such effects have been reported in the past,  HFS is still often neglected in modern spectroscopic studies. \\

\item { \it Microturbulence}: We derived abundances changing the broadening parameter \vmic. We found that all methods have { the same dependency of the derived abundances on \vmic}, in which the strongest lines have the strongest dependencies.  { In the most extreme case},  the final abundances can differ by 1.2~dex when \vmic\ has values of 1~km/s different.  { Typical differences of 0.2~km/s correspond to differences of 0.1~dex in abundance for strong lines ($\log {\mathrm{EW}/\lambda} \sim -4.8$) and of 0.05~dex for weaker lines.}\\

\item {\it $\alpha$-element abundances and continuum opacities}: Differences of less than 0.1~dex are obtained when scaling or not the  abundances with the corresponding $\alpha-$enhancement of the atmosphere model. Cool stars are more affected than metal-poor stars, reflecting how $\alpha-$abundances affect the overall continuum opacities.   \\

\item {\it Atmospheric model interpolation}: We compared the interpolated atmosphere models by the different methods and found differences of up to 10\% in the gas pressure at $\tau_{\mathrm{ross}} = 1$ for 61~Cyg~A. We derived abundances of all elements taking two different interpolated models, and found that these differences are translated to differences in the abundances of less than 0.02~dex in most of the cases.  These differences depend on the method employed.\\

\item {\it Blends}: We investigated the impact of abundances derived from two Ca lines that have different levels of small blends. We found no significant differences in the final abundances. \\

\item {\it Radiative transfer codes}: We investigated the differences of radiative transfer codes by (1) fixing the EWs and (2) fixing the masks and convolution parameters when synthesis methods are used to derive abundances. We found maximal differences of { 0.06}~dex  for weak lines in the first case, maximal differences of { 0.12}~dex for cool stars in the second case. 
\end{enumerate}

\noindent Based on our analysis we argue that if independent pipelines are combined to provide a unique set of final results of abundances, then a fixed value for the continuum flux, as well as for the \vmic\ parameter, { should} be adopted. The derivation of these { input data} must be properly documented and  provided to the community for better reproducibility of results. Furthermore, since it is common that observed and theoretical lines can have slight differences in the central wavelength, synthesis methods { should} take this into account by shifting the observed lines to a local rest frame.

Thanks to the tests performed for this workshop we were able to improve two technical aspects in the  spectral analysis of the GBS and the GES: We improved the line \ion{Mn}{I} $\lambda 6021$~\AA\ in the GES line list, which will be adopted in its future version. In addition, the GES/GBS methods \ucm\ and \porto\ improved with the new functionality of treating HFS, which will be adopted in future  GES data releases. { In this work we provide quantitative estimates of uncertainties in elemental abundances due to typical differences in technical assumptions in spectrum modelling. These may serve as a guideline when estimating uncertainties for pipelines based on a single method.} 

 The Gaia-ESO Survey homogenises results from multiple analysis teams, in which these types of issues have been startlingly revealed. Surveys currently under development, as much based on lessons learnt from Gaia-ESO, are inclined to use preferentially the results from only one analysis pipeline \citep[see overview of e.g.][]{2016arXiv160201115A}, thereby sidestepping the issues explored here.   However, it is clear that combining surveys onto the much anticipated Gaia standard will yield the most comprehensive database of kinematic and chemical measurements of the Milky Way stellar populations to-date \citep{2015arXiv150608642F, 2016arXiv160207702B}. It is at this inter-survey level that these issues will return in full force, and therefore it is essential that we work to understand and resolve such issues now, or risk rendering these impressive datasets eternally and unsatisfactorily disconnected. 

\begin{acknowledgements}
This work resulted from a workshop organised in the Institute of Astronomy, University of Cambridge in February 2016 and was partly supported by the European Union FP7 programme through ERC grant number 320360. We thank the referee for valuable comments on the manuscript. P.J. acknowledges partial support of King's College Cambridge. U.H. acknowledges support from the Swedish National Space Board (SNSB/Rymdstyrelsen). SBC want to thank Laurent Eyer from the University of Geneva for his support for this work. V.A.  acknowledges the support of the Funda\c{c}\~ao para a Ci\^encia e Tecnologia (FCT - Portugal) in the form of the grant SFRH/BPD/70574/2010. V.A. was also supported by FCT through the research grants (ref. PTDC/FIS-AST/7073/2014 and ref. PTDC/FIS-AST/1526/2014) through national funds and by FEDER through COMPETE2020 (ref. POCI-01-0145-FEDER-016880 and ref. POCI-01-0145-FEDER-016886). This work has used ESO archive data  of the star  HD22879 taken with Programme ID: 080.D-0347(A) (PI: U. Heiter). SB acknowledges funds from the Alexander von Humboldt Foundation in the framework of the Sofja Kovalevskaja Award endowed by the Federal Ministry of Education. LC acknowledges the support by the MINECO (Spanish Ministry of Economy) - FEDER through grant ESP2014-55996-C2-1-R and MDM-2014-0369 of ICCUB (Unidad de Excelencia 'Mar\'ia de Maeztu'). LC also acknowledges financial support from the University of Barcelona under the APIF grant.
\end{acknowledgements}

\small
\bibliographystyle{aa}
\bibliography{benchmark}

\end{document}